\documentclass[a4paper,10pt]{llncs}

\usepackage{mathtools}
\usepackage{newtxmath} 
\usepackage{newtxtext}

\usepackage{latexsym}
\usepackage{amsmath}
\usepackage[linesnumbered,ruled,noend,noline,noresetcount]{algorithm2e}
\usepackage{amssymb}
\usepackage{paralist}
\usepackage{graphicx}
\usepackage{trimclip}
\usepackage{wrapfig}
\usepackage{url}
\usepackage{stmaryrd}
\usepackage{amsbsy}
\usepackage{bbm}
\usepackage{tabularx}
\usepackage{booktabs}

\usepackage{tikz}
\usetikzlibrary{arrows}
\usetikzlibrary{automata}
\usetikzlibrary{positioning,fit}
\usepackage{pgfplots}

\definecolor{darkgreen}{rgb}{0,0.6,0}
\definecolor{lightred}{rgb}{0.9,0.2,0.2}
\definecolor{grey}{rgb}{0.7,0.7,0.7}

\newcommand{\ol}[1]{\textcolor{blue}{\ifmmode \text{[OL: #1]}\else [OL: #1] \fi}}

\usepackage{microtype}

\tikzset{>=stealth'}

\InputIfFileExists{synctex.tex}{%
 \synctex=1
}{%
 \synctex=0
}

\makeatletter
\DeclareRobustCommand{\shortto}{%
  \mathrel{\mathpalette\short@to\relax}%
}

\DeclareRobustCommand{\shortminus}{%
  \mathrel{\mathpalette\short@minus\relax}%
}

\newcommand{\short@to}[2]{%
  \mkern2mu
  \clipbox{{.5\width} 0 0 0}{$\m@th#1\vphantom{+}{\rightarrow}$}%
}

\newcommand{\short@minus}[2]{%
  \mkern2mu
  \clipbox{{.5\width} 0 0 0}{$\m@th#1\vphantom{+}{-}$}%
}
\makeatother

\newcommand{\labeledto}[1]{{{\shortminus}\hspace{-2pt}\raisebox{0.16ex}{$\scriptstyle\{ #1\hspace{-0.28pt}\}$}\hspace{-2.2pt}{\shortto}}}
\newcommand{\scriptlabeledto}[1]{{{\shortminus}\hspace{-1.0pt}\raisebox{0.12ex}{$\scriptscriptstyle\{ #1\hspace{-0.28pt}\}$}\hspace{-1.6pt}{\shortto}}}
\def\move(#1,#2,#3){
\mathchoice
{#1\,\labeledto{#2}\,#3}
{#1\labeledto{#2}#3}
{#1\scriptlabeledto{#2}#3}
{#1\scriptlabeledto{#2}#3}
}
\newcommand\movestd[3]{
\mathchoice
{#1\,\labeledto{#2}\,#3}
{#1\labeledto{#2}#3}
{#1\scriptlabeledto{#2}#3}
{#1\scriptlabeledto{#2}#3}
}

\definecolor{ballblue}{rgb}{0.13, 0.67, 0.8}
\newcommand{\lt}[1]{\textcolor{ballblue}{\ifmmode \text{[LT: #1]}\else [LT: #1] \fi}}
\newcommand{\td}[1]{\textcolor{blue}{\ifmmode \text{[#1]}\else [#1] \fi}}
\newcommand{\tv}[1]{\textcolor{magenta}{\ifmmode \text{[TV: #1]}\else [TV: #1] \fi}}
\newcommand{\lh}[1]{\textcolor{orange}{\ifmmode \text{[LH: #1]}\else [LH: #1] \fi}}
\newcommand{\mv}[1]{\textcolor{purple}{\ifmmode \text{[MV: #1]}\else [MV: #1] \fi}}
\newcommand{\os}[1]{\textcolor{cyan}{\ifmmode \text{[OS: #1]}\else [OS: #1] \fi}}

\newcommand{\blinded}[1]{\ifx\blindreview\undefined #1 \else \textcolor{black!65}{[blinded for review]}\fi}

\newcommand{\nat}{\mathbb{N}}

\newcommand{\secref}[1]{\S{\ref{#1}}}

\newcommand{\system}{T} 
\newcommand{\var}{x} 
\newcommand{\vars}{V} 
\newcommand{\svar}{\texttt{l}} 
\newcommand{\symbols}{\Sigma} 
\newcommand{\symb}{a} 
\newcommand{\asymb}{\texttt{a}} 
\newcommand{\conf}{\alpha} 
\newcommand{\confs}{\mathcal{C}} 
\newcommand{\assignments}{\mathit{f}} 
\newcommand{\psphere}{\psi} 
\newcommand{\semof}[1]{\llbracket#1\rrbracket} 
\newcommand{\lang}[0]{\mathcal{L}} 
\newcommand{\langof}[1]{\lang(#1)} 

\newcommand{\aut}{A} 
\newcommand{\states}{Q} 
\newcommand{\state}{q}
\newcommand{\stateone}{q}
\newcommand{\statetwo}{r}
\newcommand{\stvar}{\texttt{s}} 
\newcommand{\trel}{\Delta} 
\newcommand{\cvar}{c} 
\newcommand{\cvars}{C} 
\newcommand{\sguard}{\sigma} 
\newcommand{\guard}{g} 
\newcommand{\update}{\mathit{u}} 
\tikzset{mstatestyle/.style={draw,rectangle,rounded corners,fill=blue!0,inner xsep=.3em,inner ysep=0em,text height=2ex,text depth=.9ex}}
\newcommand{\mstate}[1]{\left(#1\right)} 

\newcommand{\subst}{\theta} 
\newcommand{\substc}{\subst_{\mathit{const}}} 
\newcommand{\substa}{\subst_{\mathit{at}}} 
\newcommand{\unprime}{\subst_{\mathit{unprime}}} 
\newcommand{\rename}{\subst_{\mathit{rename}}} 
\newcommand{\renamePrimed}{\subst'_{\mathit{rename}}} 
\newcommand{\domof}[1]{\mathit{dom(#1)}} 
\newcommand{\Pars}{\mathcal{P}} 
\newcommand{\subs}[1]{[#1]} 
\newcommand{\elimof}[2]{{\mathrlap{\mathrlap{\hspace*{0.7mm}\exists}{\hspace*{0.35mm}\exists}}{\exists\hspace{0.8mm}}} #1:#2} 
\newcommand{\issat}[0]{\texttt{sat}}
\newcommand{\issatof}[1]{\issat(#1)}

\newcommand{\defeq}[0]{\mathrel{\stackrel{\textrm{\tiny def}}{=}}}
\newcommand{\defequiv}{\defeq}

\newcommand{\wl}{\mathit{Worklist}} 
\newcommand{\ass}{\leftarrow} 
\newcommand{\processed}{\psphere} 

\newcommand{\Max}{\textit{\textbf{max}}}

\renewcommand{\max}{\Max}

\newcommand{\emp}{\varepsilon}

\newcommand\termsof[1]{\mathit{terms}(#1)}
\newcommand\substf[1]{\varphi_{#1}}
\newcommand\init{I}

\newcommand\outcome{\mathit{out}}
\newcommand\successor[3]{#1\xrightarrow{#2}#3}
\newcommand\num[1]{\mathit{num}(#1)}
\newcommand\at[1]{\mathit{at}(#1)}

\newcommand{\detaut}[0]{\aut^{d}}

\newcommand{\fsphere}{\psphere}
\newcommand{\minterm}{\mu}
\newcommand{\True}{\top}
\newcommand{\False}{\bot}

\newcommand{\mean}{\mu}
\newcommand{\median}{m}
\newcommand{\stdev}{\sigma}

\newcommand{\dist}[0]{\mathit{dist}}
\newcommand{\distof}[1]{\dist(#1)}
\newcommand{\Pone}{p_0}
\newcommand{\Ptwo}{p_1}
\newcommand{\Pthree}{p_2}

\author{
  Luk\'{a}\v{s} Hol\'{i}k\inst{1},
  Ond\v{r}ej Leng\'{a}l\inst{1}, 
  Olli Saarikivi\inst{2},\\
  Lenka Turo\v nov\'a \inst{1},
  Margus Veanes\inst{2},
  Tom\'{a}\v{s} Vojnar\inst{1}
}

\institute{
  {FIT, Brno University of Technology, IT4Innovations Centre of Excellence,
   Czech~Republic}
  \and
  {Microsoft Research, Redmond, USA}
}

\title{
Succinct Determinisation of Counting Automata via Sphere Construction (Technical Report)
\thanks{
This work has been supported by the Czech Science Foundation (project No. 19-24397S), the IT4Innovations Excellence in Science (project No. LQ1602), and the FIT BUT internal project FIT-S-17-4014.
}
}

\begin{document} 

\maketitle
\thispagestyle{empty}

\vspace*{-2mm}
\begin{abstract} We propose an efficient algorithm for determinising counting automata
(CAs), i.e., finite automata extended with bounded counters.
The algorithm avoids unfolding counters into control states, unlike the na\"{\i}ve approach, 
and thus produces much smaller deterministic automata. 
We also develop a simplified and faster version of the general algorithm for the
  sub-class of so-called monadic CAs (MCAs), i.e., CAs
with counting loops on character classes, which are common in practice.
Our main motivation is (besides applications in verification and decision procedures of logics) 
the application of deterministic (M)CAs in pattern matching regular expressions 
with counting, which are very common in e.g. 
network traffic processing and log analysis.
We have evaluated our algorithm against practical benchmarks from these application domains 
and concluded that compared to the na\"{\i}ve approach, 
our algorithm is much less prone to explode, 
produces automata that can be several orders of magnitude smaller, 
and is overall faster. 
\end{abstract}

\vspace*{-5.0mm}
\section{Introduction}\label{sec:intro}
\vspace{-1.0mm}

The \emph{counting operator}---also known as the operator of \emph{limited
repetition}---is an operator commonly used in extended regular
expressions (also called \emph{regexes}).
Limited repetitions do not extend expressiveness beyond regularity,
but allow one to succinctly express patterns such as \texttt{(ab)\{1,100\}}
representing all words where \texttt{ab} appears 1--100 times.
Such expressions are very common (cf.~\cite{cikm15}), e.g., in the RegExLib
library \cite{regexlib}, which collects expressions for recognising URIs, markup
code, pieces of Java code, or SQL queries; in the Snort rules \cite{snort} used
for finding attacks in network traffic; or in real-life XML schemas, with the
counter bounds being as large as 10 million \cite{cikm15}.
This observation is confirmed by our own experiments with patterns provided by
Microsoft for verifying absence of information leakage from network traffic
logs.
%
%
Counting constraints may also naturally arise in other contexts, such as in
automata-based verification approaches (e.g.~\cite{ultimate:ketchup})
for describing sets of runs through a loop with some number of
repetitions.

Several finite automata counterparts of regular counting constraints have
appeared in the literature (e.g.~\cite{Hovland09,KilTuh07,jha_extended,Sperberg-McQueen-ExtendedFAWeb}),
all essentially boiling down to variations on counter automata with counters limited to a bounded range of values. 
Such counters do not extend the expressive power beyond regularity, but bring succinctness, exactly as the counters in extended regular expressions.
In this paper, we call these automata \mbox{\emph{counting automata} (CAs).}

The main contribution of this paper is a novel \emph{succinct determinisation}
of CAs.
Our main motivation is in \emph{pattern matching}, where deterministic automata
allow for algorithms running reliably in time linear to the length of the text.
However, the na\"{\i}ve determinisation of CAs (and counting constraints in
general)---which encodes counter values as parts of control states, leading to
classical nondeterministic finite automata (NFAs), which are then determinised
using the standard subset construction---can easily lead to state explosion,
causing the approach to fail.
See, e.g., the CA in Fig.~\ref{fig:littleA}, for which the minimal deterministic
finite automaton (DFA) has $2^{k+1}$ states with $k$ being the upper bound of
the counter.
\emph{Backtracking}-based algorithms, which can be used instead, are slower and
unpredictable, may easily require a prohibitively large time, and are even prone
to DoS attacks, cf. \cite{owasp}.
%
%
A viable alternative is \emph{on-the-fly determinisation}, which determinises
only the part of the given NFA through which
the input word passes, as proposed already in~\cite{Thom68}.
However, the overhead during matching might be significant,  
and the construction can still explode on some words, much like the full
determinisation, especially when large bounds on counters are used (which, in
our experience, makes some regex matchers to give up already the translation to
NFAs).

Our algorithm, which allows one to \emph{succinctly} determinise CAs, is
therefore a~major step towards alleviating the above problems by making the
determinisation-based algorithms applicable more widely.
We note that this has been an open problem (whose importance was stressed, e.g.,
in \cite{Sperberg-McQueen-ExtendedFAWeb}) that a number of other works, such as
\cite{Hovland09,KilTuh07}, have attempted to solve, but they could only cope
with very restricted fragments or alleviate the problem only partially, yielding
solutions of limited practical applicability.

Our algorithm is general and often produces small results.
Moreover, we also propose a version specialised to counting restricted to
repetition of single characters from some \emph{character class}, called
\emph{monadic counting} here (e.g.,
\texttt{[ab]\{10\}} is monadic while \texttt{(ab)\{10\}} is not).
This class is of particular practical relevance since we discovered that most of
the regular expressions with counters used in practice are of this form.
Our specialised algorithm can produce deterministic
CAs exponentially more succinct than the corresponding DFAs and its worst-case
complexity is only polynomial in the maximum values of counters (in contrast to
the exponential na\"{\i}ve construction).


We have implemented the monadic CA determinisation and
evaluated it on real-life datasets of regular expressions with monadic counting.
We found that our resulting CAs can be much smaller than minimal DFAs, are less
prone to explode, and that our algorithm, though not optimised, is overall
faster than the na\"{\i}ve determinisation that unfolds counters.
%
%
We also confirmed that monadic regexes present an important subproblem, with
over 95\,\% of regexes in the explored datasets being of this type.

\paragraph{Running example.}

\begin{wrapfigure}[7]{r}{5cm}
\vspace{-11mm}
\hspace*{-3mm}
\begin{minipage}{5cm}
  \tikzset{every state/.style={minimum size=15pt}}
  \begin{tikzpicture}[shorten >=1pt,node distance=2.5cm,on grid,initial text=$\True$]
    \node[state,initial]   (q_0)                {$q$};
    \node[state,accepting,label={right:$\{c=k\}$}] (q_1) [right=of q_0] {$r$};
    \path[->] (q_0) edge                node [above] {$\svar = \asymb,c'=0$} (q_1)
                    edge [loop above]   node         {$\True$} ()
              (q_1) edge [loop above]   node         {$c<k,c'=c+1$} ();
  \end{tikzpicture}
\end{minipage}
\vspace{-3mm}
\caption{A CA for the regex \texttt{.*a.\{$k$\}} with $k \in \nat$,
  $I:\stvar=q$, $F:\stvar=r\land \cvar = k$, and $\trel : \move(q,{\True,\True},q)
  \lor \move(q,{\svar =\asymb,c'=0},r) \lor \move(r,{c<k,c'=c+1},r)$.}
\label{fig:littleA}
\end{wrapfigure}

\begin{figure}[t]
\begin{center}
\tikzset{every state/.style={minimum size=15pt}}
\begin{tikzpicture}[shorten >=1pt,on grid,initial text=$\True$]
  \tikzset{bendyr/.style={bend right=17}}
  \node[mstatestyle,initial]   (q_0)                {$\{q\mapsto 1\}$};
  \node[mstatestyle,node distance=3.3cm,accepting,label={left:$\{p_0 = 1\}$}] (q_1) [right=of q_0]
    {$\{q\mapsto1,r\mapsto1\}$};
  \node[mstatestyle,node distance=6cm,accepting,label={left:$\{p_1 = 1\}$}] (q_2) [right=of q_1]
    {$\{q\mapsto 1,r\mapsto 2\}$};

  \path[->] (q_0) edge [loop above]   node         {$\svar \neq a$} ();
  \path[->] (q_0) edge[bendyr]        node [below,align=center]
              {$\svar =a$\\$p_0' = 0$} (q_1);

  \path[->] (q_1) edge [loop above]   node[align=center]
              {$\svar \neq a$, $p_0 < 1$\\ $p_0'=p_0+1$} ();
  \path[->] (q_1) edge [loop below]   node[align=center]
              {$\svar = a$, $p_0 = 1$\\ $p_0'=0$} ();
  \path[->] (q_1) edge[bendyr]        node [below,align=center]
              {$\svar =a$, $p_0 < 1$\\ $p_0'=0$, $p_1' = p_0 + 1$} (q_2);
  \path[->] (q_1) edge[bendyr]        node [above,align=center]
              {$\svar \neq a$\\$p_0 = 1$} (q_0);

  \path[->] (q_2) edge [loop above]   node[align=center]
              {$\svar \neq a$, $p_1<1$\\$p_0'=p_0+1$, $p_1'=p_1+1$} ();
  \path[->] (q_2) edge [loop below]   node[align=center]
              {$\svar = a$, $p_1=1$\\$p_0'=0$, $p_1'=p_0+1$} ();
  \path[->] (q_2) edge[bendyr]        node [above,align=center]
              {$\svar \neq a$, $p_1 = 1$\\ $p_0'=p_0+1$} (q_1);
\end{tikzpicture}
\end{center}
\vspace{-0.8cm}
\caption{The DCA generated from the CA of Fig.~\ref{fig:littleA} for $k=1$ by our algorithm for determinisation of monadic CA (Section~\ref{sec:monadicalgo}).}
\label{fig:littleAdet}
\end{figure}
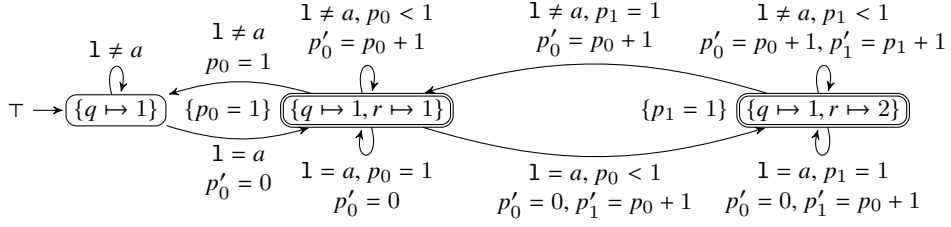

To illustrate our algorithms, consider the regex $\texttt{.*a.\{$k$\}}$ where
$k\in\nat$.
It says that the $(k+1)$-th letter from the end of the word must be~\asymb.
The minimal DFA accepting the language has $2^{k+1}$ states since it must
remember in its control states the positions of all letters~\asymb{} that were
seen during the last $k+1$ steps. 
For this, it needs a finite memory of $k+1$ bits, which has $2^{k+1}$ reachable
configurations.
The regex corresponds to the nondeterministic CA of Fig.~\ref{fig:littleA}.
In the transition labels, the predicates over the variable~$\svar$ constrain the
input symbol, the predicates over~$c$ constrain the current value of the
counter~$c$, and the primed variant of~$c$, i.e., $c'$, stands for the value
of~$c$ after taking the transition.  The initial value of~$c$ is unrestricted,
and the automaton accepts in the state~$r$ if the value of~$c$ equals~$k$.
Our monadic determinisation algorithm, presented in
Section~\ref{sec:monadicalgo}, then outputs the deterministic CA (DCA) of
Fig.~\ref{fig:littleAdet} (for $k=1$).
Intuitively, it uses $k+1$ counters to remember how far back the last $k+1$
occurrences of $a$ appeared.
Depending on $k$, the resulting DCA has $k+2$ states, $4(k+1)+1$ transitions,
and $k+1$ counters.
That is, its size is linear to $k$ in contrast to the factor $2^k$ in the size
of the~minimal DFA.
\qed

\section{Counting Automata}\label{sec:prelims}
\vspace{-2.0mm}

\paragraph{Preliminaries.}
We 
use~$\nat$ to denote the set of natural numbers $\{0,1,2,\ldots\}$.
Given a~function $f: A \to B$, we refer to the elements of~$f$ using $a \mapsto
b$ (when $f(a) = b$).
For the rest of the paper, we consider a fixed finite \emph{alphabet} $\symbols$ of \emph{symbols}. 
A~word over~$\symbols$ is a~finite sequence of symbols $w = a_1 \cdots a_n \in \symbols^*$.
We use $\epsilon$ to denote the \emph{empty word}.

Given a~set of variables~$\vars$ and a~set of constants~$\states$ (disjoint with $\nat$), 
we define a~\emph{$\states$-formula over~$\vars$} to be a~quantifier-free
formula~$\varphi$ of Presburger arithmetic extended with
constants from~$\states$ and~$\symbols$,
i.e., a~Boolean combination
of (in-)equalities~$t_1 = t_2$ or $t_1 \leq t_2$ where~$t_1$ and~$t_2$ are
constructed using~$+$, $\nat$, and~$\vars$, and predicates of the form $\var =
a$ or $\var = q$ for $\var \in \vars$, $a\in\symbols$, and $q\in\states$.
An assignment $M$ to free variables of~$\varphi$ is a \emph{model} of $\varphi$,
denoted as $M \models \varphi$, if it makes~$\varphi$ true.
We use $\issatof \varphi$ to \mbox{denote that $\varphi$ has a~model.}

Given a formula $\varphi$ and a (partial) map $\subst:\termsof{\varphi} \to S$, where
$\termsof\varphi$ denotes the set of terms in~$\varphi$ and~$S$ is some set of
terms,
$\varphi \subs\subst$ denotes a~\emph{term substitution}, i.e., the
formula~$\varphi$ with all occurrences of every term $t\in\domof\subst$ replaced by
$\subst(t)$. As usual, replacing a~larger term takes priority over replacing
its subterms (we treat primed variables and parameters as atomic terms, hence $(p'=
1)\subs{\{p\mapsto q\} }$ is still $p' = 1$).
The \emph{substitution formula}~$\substf\subst$ of $\subst$ is defined as the conjunction of equalities
\mbox{
$
\substf\subst \defequiv \bigwedge_{t\in \domof\subst} (\subst(t) = t)
$.}
Finally, the set of \emph{minterms} of a~finite set $\Phi$ of predicates
is defined as the set of all satisfiable predicates of
$\{\bigwedge_{\phi\in \Phi'} \phi \land \bigwedge_{\phi\in\Phi\setminus \Phi'}\neg \phi  \mid \Phi'\subseteq \Phi\}$.

\paragraph{Labelled transition systems.}

\vspace{-2mm} We will introduce our counting automata, such as that of
Fig.\ref{fig:littleA}, as a specialisation of the more general model of labelled
transition systems.
This perspective and related notation allows for a more abstract and concise
formulation of our algorithms than the more standard approach, in which one
would define counting automata in a more straightforward manner as an extension
of the classical finite automata.

A \emph{labelled transition system} (LTS) over~$\symbols$ is a tuple
$\system = (\states,\vars,I,F,\trel)$ where
$\states$ is a finite set of \emph{control states}, 
$\vars$ is a finite set of \emph{configuration variables}, 
$I$ is the \emph{initial $\states$-formula} over~$\vars$, 
$F$ is the \emph{final $\states$-formula} over~$\vars$, 
and $\trel$ is the \emph{transition $\states$-formula} over $\vars \cup \vars' \cup \{\svar\}$
with $\vars' = \{\var'\mid \var\in \vars\}$, $\vars \cap \vars' = \emptyset$, and
$\svar \not \in \vars$.
We call $\svar$ the \emph{letter/symbol variable} and allow it as the only term
that can occur within a~predicate $\svar = a$ for $a\in\symbols$, called an \emph{atomic symbol guard}.%
\footnote{To handle large or infinite sets of symbols symbolically, the predicates $\svar = a$ may be generalised to predicates from an arbitrary effective Boolean algebra, as in \cite{SfaMinPOPL14}.}
Moreover, $\svar$ is also not allowed to occur in any other predicates in~$\trel$.
A~\emph{configuration} is an assignment $\conf:
\vars\rightarrow\nat\cup\states$ that maps
every configuration variable to a~number from~$\nat$ or a~state from
$\states$.
Let $\confs$ be the set of all configurations.
The transition formula~$\trel$ encodes the transition relation
$\semof\trel \subseteq \confs\times\symbols\times\confs$ such that 
$(\conf,\symb,\conf')\in\semof\trel$ iff $\conf\cup\{\var'\mapsto k\mid \conf'(\var)=k\} \cup \{\svar\mapsto\symb\} \models\trel$.
We use $|\trel|$ to denote the size of~$\semof \trel$.
For a~word $w \in \symbols^*$, we define inductively that
a~configuration~$\conf'$ is a~\emph{$w$-successor} of~$\conf$, 
written $\successor{\conf}{w}{\conf'}$,
such that $\successor{\conf}{\epsilon}{\conf}$ for all $\conf\in\confs$, and
$\successor{\conf}{\symb v}\conf'$ iff
$\successor{\conf}{\symb}{\bar\conf}\successor{}{v}{\conf'}$ for some
$\bar\conf \in \confs$, $\symb\in \symbols$, and $v \in \symbols^*$.
A~configuration~$\conf$ is \emph{initial} or \emph{final} if $\conf\models I$ or $\conf\models F$, respectively.
The \emph{outcome} of $\system$ on a~word~$w$ is the set $\outcome_\system(w)$ of all
$w$-successors of the initial configurations, and $w$ is accepted by~$\system$ if
$\outcome_{\system}(w)$ contains a final configuration.
The \emph{language} $\langof \system$ of $\system$ is \mbox{the set of all words that
$\system$~accepts.}

\vspace{-2mm}
\paragraph{Counting automata.}

A \emph{counting variable (counter)} is a configuration variable $\cvar$ whose
value ranges over~$\nat$ and which can appear (within
$\trel$, $I$, and $F$) only in \emph{atomic counter guards} of the form $\cvar\leq
k,\cvar\geq k$, (using $<,=,>$ as syntactic sugar) or \emph{term equality tests} $t_1 =
t_2$, and in \emph{atomic counter assignments} $\cvar' = t$ with $t,t_1,t_2$ being \emph{arithmetic
terms} of the form $d + k$ or $k$ with $k\in\nat$ and $d$ being a~counter.
A \emph{control state variable} is a~variable~$\stvar$ whose value ranges over
states $\states$ and appears only in \emph{atomic state guards}
$\stvar = \state$ and \emph{atomic state assignments} $\stvar'= \state$ for $\state\in \states$.
A Boolean combination of atomic guards (counter, state, or symbol) is a \emph{guard formula} and a Boolean combination of atomic assignments is an \emph{assignment formula}.

A \emph{(nondeterministic) counting automaton} (CA) is a~tuple $\aut =
(\states, \cvars,I,F,\trel)$ 
such that 
$(\states,\vars, I, F, \trel)$ is an LTS with the following properties:
(1)
The set of configuration variables 
$\vars = \cvars\cup\{\stvar\}$ consists of a set of counters $\cvars$ and a single control state variable~$\stvar$ s.t.~$\stvar\not\in\cvars$. 
(2)~The transition formula $\trel$ is a disjunction of \emph{transitions}, which
are conjunctions of the form $\stvar = \stateone \land \guard\land\assignments
\land \stvar' = \statetwo$, denoted by
$\move(\stateone,{\guard,\assignments},\statetwo)$, where $q, r \in \states$, $\guard$
is the transition's~\emph{guard formula} over $\vars\cup\{\svar\}$, and $\assignments$ is the transition's
\emph{counter assignment formula}, a conjunction of atomic assignments to counters,
in which every counter is assigned at most once.
(3)~There is a constant $\max_\aut \in \nat$ such that no counter can ever grow
above that value, i.e., $\forall c \in C ~\forall w \in \Sigma^* ~\forall \conf
\in \outcome_\system(w): \conf \models c \leq \max_\aut$.

The last condition in the definition of CAs is semantic and can be achieved in
different ways in practice.
For instance, regular expressions can be compiled to CAs where assignment terms
are of the form $c+1$, $0$, or $c$ only, and every appearance of $c+1$ is paired with a
guard containing a constraint $c \leq k$ for some $k \in \nat$.
In this case, $\max_\aut = K+1$ where $K$ is the maximum constant used in the
guards of the form $c \leq k$.

We will often consider the initial and final formulae of CAs given
as a~disjunction $\bigvee_{\state\in \states}(\stvar = \state \land
\varphi_\state)$ where $\varphi_\state$ is a~formula over counter guards,
in which case we write $I(\state)$ or $F(\state)$ to denote the disjunct
$\varphi_\state$ of the initial or final formula, respectively.
An example of a CA is given
in Fig.~\ref{fig:littleA}.

A \emph{deterministic counting automaton} (DCA) is a CA $\aut$ 
where $I$ has at most one model and, for every symbol $a\in\symbols$, every
reachable configuration $\conf$ has at most one $a$-successor 
(equivalently, the outcome of every word in $\aut$ is either a~singleton or the
empty set).
Finally, in the special case when $\cvars = \emptyset$, the CA is a (classical)
nondeterministic \emph{finite automaton} (NFA), or a deterministic finite
automaton (DFA) if it is deterministic. 


\vspace{-3.0mm}
\section{Determinisation of Counting Automata}\label{sec:deter}
\vspace{-2.0mm}

In this section, we discuss an algorithm for determinising CAs.
A~na\"{\i}ve determinisation converts a~given CA~$A$ into an NFA by hard-wiring
counter configurations as a~part of control states, followed by the
classical subset construction to determinise the obtained NFA
(the NFA is finite due to the bounds on the maximum values of counters).
The state space of the obtained DFA then consists of all reachable outcomes
of~$A$. 
By determinising~$A$ in this way, the succinctness of using counters is lost,
and the size of the DFA can explode exponentially not only in the number of
control states of~$A$ but also in the number of reachable counter valuations,
which makes the construction impractical.
Instead, our construction will retain counters (though their number may grow)
and represent possible word outcomes as configurations of the resulting DCA.

\vspace{-2mm}
\paragraph{Spheres.}
In particular, the outcome of a~word~$w \in \symbols^*$ in a~CA $\aut = (\states, \cvars, I, F, \Delta)$ can be
represented as a~formula~$\varphi$ over equalities of the form $\cvar = k$ and
$\stvar = \state$ where $\state\in\states$, $\cvar\in \cvars$, $k \in \nat$.
Intuitively, disjunctions can be used to obtain a single formula for the
possibly many configurations reachable in $\aut$ over $w$.
For example, the outcome of the word \texttt{aab} in Fig.~\ref{fig:littleA} is
$\varphi : \stvar = q \lor (\stvar = r \land (c = 1 \lor c = 2))$. 
Generally, the outcome of $\texttt{aab}^i$, for $0\leq i < k$, assuming $k > 2$,  is 
$\varphi_i : \stvar = q \lor (\stvar = r \land (c = i \lor c = i+1))$.

A crucial notion for our construction is then the notion of \emph{sphere}.
A sphere $\psphere$ arises from an outcome $\varphi$ by replacing the constants from
$\nat$ by parameters drawn from a~countable set $\Pars$ (disjoint from
$\nat$, $\vars$, $\states$, and $\{\svar,\stvar\}$). 
%
In the example above, the sphere obtained from the $\varphi$ is
$\psphere : \stvar = q \lor (\stvar = r \land (c = p_0 \lor c = p_1))$, and the same sphere arises from all outcomes $\varphi_i$ with $0\leq i < k$.

Spheres will play the role of the control states of the resulting DCA.
The idea of the construction is that the outcome of every word $w$ in a~DCA
$\aut^d$ will 
contain a single configuration ($\aut^d$ is deterministic) consisting of a sphere $\psphere$ as the control state and a~valuation of its parameters $\eta:\Pars\rightarrow \nat$.
The construction will ensure that $\psphere \subs{\eta}$ models the outcome $\outcome_\aut(w)$ of $w$ in $\aut$. 
In our example, the outcome of \texttt{aab} in $\aut^d$ would contain the single configuration
$\{\stvar\mapsto\psphere,p_0\mapsto 1,p_1\mapsto 2\}$, 
and the outcome of each $\varphi_i$, for $0\leq i < k$, would contain the single configuration 
$\{\stvar\mapsto\psphere,p_0\mapsto i,p_1\mapsto i+1\}$.
The example shows the advantage of our construction. 
Every outcome $\varphi_i$ would be a~control state of the na\"{i}vely determinised automaton, 
with a $b$-transition from each $\varphi_j$ to $\varphi_{j+1}$, for $0\leq j < k-1$.
In contrast to that, all these states and transitions will be in $\aut^d$ replaced by a single control state $\psphere$
with a single $b$-labelled self-loop that increments both~$p_0$ and~$p_1$.
This structure can be seen in Fig.~\ref{fig:littleAdet} (states are spheres,
labelled by their multiset representation introduced in Section~\ref{sec:monadicalgo}).

\subsection{Determinisation by Sphere Construction}


We now provide a basic version of our sphere-based determinisation, which can
also be viewed as an algorithm that constructs parametric versions of the
subsets used in subset-based determinisation.
For this basic algorithm, termination is not guaranteed, but it serves as
a~basis on which we will subsequently build a~terminating algorithm. 
Let us first introduce some needed additional notation. 

Given a formula $\varphi$, we denote by $\at\varphi$ and by $\num\varphi$ the sets of
assignment terms and numerical constants, respectively, appearing in $\varphi$.
We will use the set $\Pars' = \{p'\mid p \in \Pars\}$ and the substitution
$\unprime = \{p' \mapsto p\mid p\in\Pars\}$. 
We say that a formula over variables $\vars\cup\vars'\cup\{\svar\}\cup \Pars$ is
\emph{factorised wrt guards} if it is a~disjunction $\bigvee_{i=1}^n
(\guard_i)\land(\update_i)$ of factors, each consisting of a guard $\guard_i$
over $\vars\cup\{\svar\}\cup \Pars$ and an update formula $\update_i$ over atomic assignments
such that  
the guards of any two different
factors are mutually exclusive, i.e., $\guard_i \land \guard_j$ is unsatisfiable
for any $1\leq i\neq j\leq n$.\footnote{
A Boolean combination of atomic guards and updates can be factorised through 
(1) a transformation to DNF, yielding a set of clauses $X$; 
(2) writing each clause $\varphi\in X$ as a conjunction of a guard formula $g_\varphi$ and an assignment formula $f_\varphi$; 
(3) computing minterms of the set $\{g_\varphi\mid \varphi\in X\}$;
(4) creating one factor $(g)\land(f)$ from every minterm $g$ where $f$ is the
disjunction of all the assignment formulae $f_\varphi$ with $\varphi\in X$ compatible with $g$ (i.e., such that $g\land f_\varphi$ is satisfiable).
}
For 
a~set of variables~$U$, we denote by $\elimof{U}
\varphi$ a~formula obtained by eliminating all variables in~$U$
from~$\varphi$ (i.e., a~quantifier-free formula equivalent to $\exists
U: \varphi$).\footnote{
We note that we only need to use a specialised, simple, and cheap quantifier elimination.
In particular, we only need to eliminate counter variables $c$ from formulae such that, in clauses of their DNF, 
$c$ always appears together with a predicate $c=p$ where $p$ is a parameter. 
Eliminating $c$ from such a DNF clause is then done by simply substituting occurrences of $c$ by $p$. 
We do not need complex algorithms such as the general quantifier elimination for Presburger arithmetic. }

\newcommand{\ground}{\mathit{ground}}

\begin{figure}[t]
\begin{center}
\begin{algorithm}[H]
\caption{Sphere-based CA determinisation (non-terminating)}
\label{algo:basic}
\KwIn{A CA $\aut = (\states,\cvars,I,F,\trel)$.}
\KwOut{A DCA $\detaut = (\states^d,P,I^d,F^d,\trel^d)$ s.t. 
  $\langof \aut = \langof{\aut^d}$.}
	$\states^d\ass\wl\ass\emptyset;\,\trel^d\ass\False$\;
	$\psphere_I \ass I \subs\substc$ for some total injection 
          $\substc: \num{\init}\rightarrow \Pars$\;\label{ln:substc}
	$I^d \ass \stvar = \psphere_I\land \substf{\substc}$\;\label{ln:initf}
	add $\psphere_I$ to $\states^d$ and to $\wl$\;\label{ln:initstate}
	\While{$\wl \neq \emptyset$}{\label{ln:repeat}
		$\processed \ass \mathit{pop}(\wl)$\;\label{ln:pop}
		Let $\bigvee_{i = 1}^n (\guard_i)\land(\update_i)$ be the formula
                  $\elimof{\cvars,\stvar}{\processed\land\trel}$ factorised wrt 
                  guards\;\label{ln:factorise}
		\ForEach{$1\leq i \leq n$}{\label{ln:for}
			$\psphere_i \ass \update_i \subs\substa\subs\unprime$ 
                          for a total injection $\substa:\at{\update_i}\rightarrow
                          \Pars'$\;\label{ln:substa}
			add $\move(\processed,{\guard_i,\substf{\substa}}, 
                          \psphere_i)$ to $\trel^d$\;\label{ln:addDelta}
			\lIf{$\psphere_i \not\in \states^d$}{
				add $\psphere_i$ to $\states^d$ and to 
                                  $\wl$\label{ln:addQ}
			} 
		}
	}
	$P\ass\text{all parameters found in }\states^d$\label{ln:P}\;
	$F^d \ass \bigvee_{\psphere\in\states^d} \stvar = \psphere \land \elimof{ \cvars,\stvar}{ \psphere \land F}$\label{ln:F}\;
  $I^d \ass \ground(I^d);\trel^d \ass \ground(\trel^d)$\label{ln:ground}\;
  \KwRet $\aut^d = (\states^d,P,I^d,F^d,\trel^d)$\;
  \label{ln:ret}
\end{algorithm}		
\end{center}
\vspace*{-7mm}
\end{figure}

\vspace*{-2mm}
\paragraph{The algorithm.}
The core of our determinisation algorithm is the sphere construction described
in Algorithm~\ref{algo:basic}.
It builds a DCA $A^d = (\states^d,P,I^d,F^d,\trel^d)$ whose control
states~$\states^d$ are spheres.
Its counters are parameters from the set $P$ that is built during the run of the
algorithm.
The initial formula $I^d$ defined on line~\ref{ln:initf} assigns to~$\stvar$ the
initial control state~$\psphere_I$ (obtained on line~\ref{ln:substc}), which is
a~parametric version of~$I$ with integer constants replaced by parameters
according to the renaming~$\substc$.
Moreover, $I^d$ also equates the parameters in~$\psphere_I$ with the constants
they are replacing in~$I$.
Hence, the formula~$\psphere_I\subs{\substc^{-1}}$ models exactly the initial
configurations of~$\aut$.

\begin{example} In the running example (Fig.~\ref{fig:littleA}),
whenever referring to some variable that is assigned multiple times during the
run of the algorithm, we use superscripts to distinguish the different
assignments during the run.
On lines 1--4, the initial
sphere $\psphere_I$ is assigned the formula $\stvar=q$, and the
initial formula $I^d$ is set to $\stvar = \psphere_I$, which specifies that
$\psphere_I$ is indeed the initial control state only ($I$ does not constrain
counters, hence $I^d$ does not talk about parameters).\qed\end{example}


The remaining states of $\states^d$ and transitions of $\trel^d$ are
computed by a worklist algorithm on line~\ref{ln:repeat} with the worklist
initialised with~$\psphere_I$.
Every iteration computes the outgoing transitions of a~control state
$\psphere\in\wl$ as follows:
On line~\ref{ln:factorise}, after eliminating~$\cvars \cup \{\stvar\}$ from the formula
$\psphere\land\trel$, which describes how the next state and
counter values depend on the input symbol and the current values of
parameters, it is transformed into a~guard-factorised form.
%

\begin{example}
\label{ex:factors}
When $\psphere_I$ is taken from $\wl$ as $\processed^1$ on line~6, 
its processing starts by factorising $\elimof{\{c,\stvar\}}{\processed^1\land\trel}$ on line 7. 
Here, $\psphere^1 \land \trel$ is the formula $\stvar = q \land \left(
\move(q,{\True,\True},q) \lor \move(q,{\svar=\asymb,c'=0},r) \lor \move(r,{c<k,c'=c+1},r)
\right)$, which can be also written as\\[-3mm] 
$$
\stvar = q \land \left(\stvar'=q \lor (\svar =\asymb \land c' = 0 \land \stvar'=r )\right).
\vspace{-1mm}
$$
The elimination of $\{c,\stvar\}$ gives the formula 
$\stvar'=q \lor (\svar =\asymb \land c' = 0 \land \stvar'=r)$.
This formula is factorised into the following two factors: 
\begin{compactitem}
\item[~~($F_1$)~~]
$(\svar = \asymb) \land (\stvar'=q \lor (c' = 0 \land \stvar'=r)),$ 
\item[~~($F_2$)~~]
$(\svar \neq \asymb) \land (\stvar'=q)$.
\qed
\end{compactitem}
\end{example}

In the for-loop on line~\ref{ln:for}, every factor $(\guard_i)\land(\update_i)$
is turned into a transition with the guard~$\guard_i$; the mutual
incompatibility of the guards guarantees determinism.
The formula~$\update_i$ describes the target sphere in terms of the parameters
of the source sphere $\processed$, updated according to the transition relation.
That is, it is a~Boolean combination of
assignments of the form $\cvar' = p + k$ or $\cvar' = k$ for
$\cvar\in\cvars,p\in\Pars$, and $k\in\nat$.
Line~\ref{ln:substa} creates a sphere by substituting each of the assignment
terms (of the form $p + k$ or~$k$) with a~parameter and replacing primed
variables by their unprimed versions.\footnote{The choice of the parameters in the image of $\substa:\at{\update_i}\rightarrow\Pars'$ on line 9 is arbitrary, although, in practice, it would be sensible to define some systematic parameter naming policy and reuse existing parameters whenever possible.}
The~corresponding assignment term substitution $\substa$ records how the values
of the new parameters are obtained from the original values of the parameters
occurring in~$\processed$.
It is used to define the assignment formula of the new transition that is added
to~$\trel^d$ on line~\ref{ln:addDelta}.
%
%
The argument justifying that the construction preserves the language is the
following:
if reading~$w \in \symbols^*$ takes $\aut^d$ to $\psphere$ with
a~parameter valuation~$\eta$
such that $\psphere\subs\eta$ is equivalent to 
$\outcome_\aut(w)$,
then reading a next symbol $a$ using a transition newly created on
line~\ref{ln:addDelta} takes $A^d$ to $\psphere'$ with the parameter valuation
$\eta'$ such that $\psphere'\subs{\eta'}$ models
$\outcome_\aut(wa)$.

\vspace{-1.5mm}
\begin{example}
Factor $F_1$ of Example~\ref{ex:factors} above is processed as follows. 
A possible choice for $\substa^1$ on line 9 is the assignment $\{0\mapsto \Pone\}$.
Its application followed by $\unprime$ creates 

\vspace{-3mm}
$$\psphere_1^1 : \stvar=q \lor (c = \Pone \land \stvar=r).
\vspace{-0.5mm}
$$
From $\substa^1$, we get the substitution formula $\substf{\substa^1} :
(\Pone' = 0)$ on line 10, and so the
transition added to $\trel^d$ is $\movestd{\mstate{\stvar = q}}{\svar = \asymb ,\Pone'
= 0}{\mstate{\stvar=q \lor (c = \Pone \land \stvar=r)}}.$
The target $\psphere_1^1$ of the transition is added to $\states^d$ and to $\wl$ on line 11.
Next, Factor $F_2$ generates the self-loop $\move(\mstate{\stvar = q},{\svar \neq
\asymb,\True},\mstate{\stvar = q})$, which ends the first iteration of the while-loop.

Let us also walk through a~part of the second iteration of the while-loop, 
in which $\psi_1^1$ is taken from $\wl$ as $\psi^2$ on line 6.
The formula $\psi^2 \land \trel$ from line 7 is $\left((\stvar = r  \land \cvar =
\Pone) \lor \stvar = q\right) \land \left( \move(q,{\True,\True},q) \lor \move(q,{\svar=\asymb,c'=0},r)
\lor \move(r,{c<k,c'=c+1},r) \right)$, which is equivalent to 
$\left(\stvar = q
\land (\stvar'=q \lor (\svar =\asymb \land c' = 0 \land \stvar'=r ))\right) \lor \bigl(\stvar =
r \land \cvar = \Pone \land c < k \land c' = c+1 \land \stvar' = r\bigr).$
The elimination of $\{c,\stvar\}$ on line 7 then gives the formula
$\bigl(\stvar'=q \lor (\svar =\asymb \land c' = 0 \land \stvar'=r)\bigr)  \lor (\Pone < k \land
c' = \Pone+1 \land \stvar' = r),$
which is factorised  into the following four factors:\begin{compactitem}

  \item[~~($F_3$)~~] $(\svar = \asymb \land \Pone < k) \land (\stvar'=q \lor (c' = 0 \land
  \stvar'=r)  \lor  (c' = \Pone+1 \land \stvar' = r))$,

  \item[~~($F_4$)~~] $(\svar \neq \asymb \land \Pone < k) \land (\stvar'=q \lor (c' = \Pone+1 \land
  \stvar' = r))$,

  \item[~~($F_5$)~~] $(\svar = \asymb \land \Pone \geq k) \land (\stvar'=q \lor (c' = 0 \land
  \stvar'=r))$, and

  \item[~~($F_6$)~~] $(\svar \neq \asymb \land \Pone \geq k) \land (\stvar'=q)$.

\end{compactitem}
In the for-loop on line~8, Factor $F_3$ is processed as follows. 
Let the chosen substitution $\substa^2$ on line 9 be
$\{\Pone+1\mapsto \Ptwo,0\mapsto \Pone\}$.
Its application followed by $\unprime$ generates\\[-2mm] 
$$\psphere_1^2 : \stvar=q \lor (c = \Pone \land \stvar = r) \lor (c = \Ptwo \land \stvar=r).$$
The substitution formula $\substf{\substa^2}$ on line 10 is
$\Ptwo' = \Pone + 1 \land \Pone' = 0$, and so $\trel^d$ \mbox{gets the new transition}
$\movestd{{\psphere_1^1}}{\svar = \asymb \land \Pone < k,\Ptwo' = \Pone + 1 \land
\Pone' = 0}{{\psphere_1^2}}.$
The evaluation of the while-loop would continue analogously.
\qed
\end{example}

In the final stage of the algorithm, when (and if) the while-loop terminates,
line~\ref{ln:P} collects the set~$P$ of all parameters used in the constructed parametric
spheres of~$\states^d$ as new counters of~$\detaut$.
Further,
line~\ref{ln:F} derives the new final formula by considering all computed
spheres, restricting them to valuations where the original final formula is
satisfied, and quantifying out the original counters.
This way, final constraints on the original counters get translated to
constraints over parameters~in~$P$.
%

\begin{example}
\label{ex:final}
In our running example, 
for the spheres discussed above, we would have 
$F(\psphere^1) : \bot$, $F(\psphere^1_1) : \Pone = 1$, and $F(\psphere_1^2) : \Pone = 1 \lor \Ptwo = 1$. 
%
\qed
\end{example}

Finally,  
line~\ref{ln:ground} applies the function $\ground$ on the initial formula
and the transition formula of the constructed automaton before returning it.
This step is needed in order to avoid nondeterminism on unused and unconstrained counters.
The function $\ground$ conjuncts constraints of the form $p = 0$ with the initial
formula and with the guard of every transition for every parameter $p \in P$ that is
so far unconstrained in the concerned formula. 
Moreover, it will introduce a reset $p' = 0$ to the assignment formula of every
transition for every counter $p \in P$ that is so far not assigned on the
concerned transition.
%
%
The 
while-loop of 
Algorithm~\ref{algo:basic} needs, however, not terminate, as
\mbox{witnessed also by our example.}
\footnote{
For this step to preserve the language of the automaton, 
we need to assume that the input CA does not assign nondeterministic values to live counters.
We are refering to the standard notion: a counter is live at a state if the value it holds at that state may influence satisfaction of some guard in the future.  
Any CA can be transformed into this form, and CAs we compile from regular expressions satisfy this condition by construction.
}

\begin{example}
\label{ex:nonterm}
Continuing in Example~\ref{ex:final}, the DCA in Fig.~\ref{fig:littleAdet} would be a part of the DCA constructed by Algorithm~\ref{algo:basic}, its states being the spheres $\psphere^1$, $\psphere_1^1$, $\psphere_1^2$ from the left, 
but the while-loop would not terminate, with $\psphere_1^2$.
Instead, it would eventually generate a successor 
of~$\psphere_1^2$, the sphere
$$\psphere_1^3:\stvar=q \lor (c = \Pone \land \stvar = r) \lor (c = \Ptwo \land
\stvar=r) \lor (c = \Pthree \land \stvar=r),
$$ 
i.e., a~sphere similar
to~$\psphere_1^2$ but extended by a~new disjunct with a~new parameter~$\Pthree$.
Repeating this, the algorithm would keep generating larger and larger spheres with more and more parameters.  
\qed
\end{example}

\vspace{-3.0mm}
\subsection{Ensuring Termination of the Sphere Construction}\label{sec:termination}

In this section, we will discus reasons for possible non-termination of
Algorithm~\ref{algo:basic} and a~way to tackle them.
The main reason is that the algorithm may generate unboundedly many parameters that correspond
to different histories of a~counter~$\cvar$ when
processing the input word (including also impossible ones in which the counter exceeds the maximum value).
The algorithm indeed ``splits'' a parameter appearing in a sphere into two
parameters in the successor sphere
when the transitions of~$A$ update the counter in two different ways.

In our terminating version of Algorithm~\ref{algo:basic}, we build on the following:
(1) distinguishing between histories that converge in the same counter value is not necessary,
they can be ``merged'', and
(2) the number of different reachable counter values is bounded (by the definition of
CAs). 
We thus enforce the
invariant of every reachable configuration of~$\detaut$ that all
parameters in the configuration have distinct values.
The invariant is enforced by testing equalities of parameters and merging
parameters with equal values on transitions of $\detaut$.
All transitions of~$\detaut$ entering spheres with
more than $\max_A +1$ parameters can then be discarded because the invariant implies that they
cannot be taken at any configuration of~$\detaut$.
Furthermore, we will also ensure that the algorithm does not diverge because of
generating semantically equivalent but syntactically different spheres (because
of different names of parameters or different formulae structure).

A terminating determinisation of CAs is obtained
from Algorithm~\ref{algo:basic} by replacing
lines~\ref{ln:substa}--\ref{ln:addQ} by the code in
Algorithm~\ref{algo:terminating}.
In order to ensure that parameters have pairwise distinct values,
the transitions of~$\detaut$ test equalities of the values assigned to parameters 
and ensure that two parameters are never used to represent the same value.
Different
histories
of counters are thus merged if they converge
into the same value.
To achieve this, Algorithm~\ref{algo:terminating} enumerates all feasible equivalences
of the assignment terms of~$\update_i$ on line~\ref{ln:ref:enumEq} and generates successor
transitions for each of them separately.
When deciding whether an equivalence $\sim$ on the assignment terms is feasible,
the algorithm performs two tests:
(1)~The formula 
$\varphi_{\sim} \defequiv \bigwedge_{t_1\sim t_2,t_1,t_2\in \at{\update_i}} (t_1 = t_2)
\land \bigwedge_{t_1\not\sim t_2,t_1,t_2\in \at{\update_i}} (t_1 \neq t_2)$ 
is tested for satisfiability, meaning that the
equivalence is not trying to merge terms that can never be equal (such as,
e.g., $p$ and~$p+1$).
(2)~The number of equivalence classes should be at most $\max_\aut +
1$ since this is the maximum number of different values that the counters can reach
due to the requirement that the values must be between $0$ and $\max_\aut$.

\begin{figure}[t]
\begin{center}
\begin{algorithm}[H]
\caption{Ensuring termination of sphere-based CA determinisation}
\label{algo:terminating}
        \ForEach{equivalence $\sim$ on $\at{\update_i}$ s.t. 
          $\issatof{\varphi_\sim}$ and $|\at{\update_i}/_{\sim}|\leq 
          \max_\aut+1$}{\label{ln:ref:enumEq}
                \KwSty{let} $\substa:\at{\update_i}\rightarrow \Pars'$ be an injection\;\label{ln:ref:substa}
                $\psphere_i \ass \update_i
                \subs{\substa}\subs{\unprime}$\;\label{ln:ref:suckessor}
                \If{$\exists \rename:\Pars\leftrightarrow\Pars ~\exists \sigma \in
                  \states^d: \psphere_i\subs{\rename} \Leftrightarrow \sigma$}{ 
                  \label{ln:ref:if}
                        add $\move(\processed,{\guard_i\land 
                          {\varphi_{\sim}}\subs{\substa},
                          \substf{\substa}\subs{\renamePrimed}}, 
                          \sigma)$ to
                          $\trel^d$\;\label{ln:ref:addDeltaOld}
                }
                \Else{
                        add $\move(\processed,{\guard_i\land 
                          {\varphi_{\sim}}\subs{\substa},\substf{\substa}}, 
                          \psphere_i)$ to $\trel^d$\;\label{ln:ref:addDelta}
                        add $\psphere_i$ to $\states^d$ and to
                          $\wl$\label{ln:ref:addQ}\;
                }
        }
\end{algorithm}		
\end{center}
\vspace*{-4mm}
\end{figure}

Line~\ref{ln:ref:substa} builds a~term assignment replacement $\substa$ that
maps all $\sim$-equivalent terms to the same
(future) parameter, and line~\ref{ln:ref:suckessor} computes the target sphere, reflecting
the given merge.
The test on line \ref{ln:ref:if} checks whether the target sphere is equal to
some already generated sphere up to a~parameter renaming (represented
by a~bijection $\rename:\Pars\leftrightarrow~\Pars$).
If so, the created sphere is discarded, and a~new transition going to the old
sphere is generated on line~\ref{ln:ref:addDeltaOld}; we need to
rename the primed parameters used in the transition's assignment appropriately
according to 
$\renamePrimed = \{\Pone'\mapsto \Ptwo'\mid \Pone \mapsto \Ptwo\in\rename\}$.
Otherwise, a transition into the new sphere is added on line
\ref{ln:ref:addDelta}, and the new sphere is added to $\states^d$ and $\wl$.
In both cases, the guard of the generated transition is extended by the formula
$\varphi_{\sim}\subs{\substa}$, which encodes the equivalence~$\sim$, and hence
explicitly enforces that~$\sim$ holds when the transition is taken.

Note that the test on the maximum number of equivalence classes can be optimised
if finer information about the maximum reachable values of the individual
counters is available.
Such information can be obtained, e.g., by looking at the constants used in the
guards of the transitions where the different counters are increased.
For any counter, one should then not generate more parameters representing its
possible values than what the upper bound on that counter is (plus one).


\begin{theorem}
Algorithm~\ref{algo:basic}
with the modification presented in Algorithm~\ref{algo:terminating} terminates
and produces a DCA with $\langof \aut = \langof
{\aut^d}$ and $|\states^d| \leq 2^{|\states|\cdot(\max_A + 1)^{|C|}}$.
\end{theorem}
\vspace{-1mm}

\begin{proof}[idea]
The fact that the algorithm indeed constructs a~DCA is because
line~\ref{ln:factorise} of Algorithm~\ref{algo:basic} generates pairwise
incompatible guards on transitions only.
It is also easy to show by induction on the length of the words that the
language is preserved.
The termination then follows from the facts that
(1)~the algorithm has a~bound on the maximum number of parameters in spheres
(ensured by the condition over~$\sim$ on line~\ref{ln:ref:enumEq} of
Algorithm~\ref{algo:terminating}) and
(2)~no spheres equal up to renaming are generated (ensured by the check on
line~\ref{ln:ref:if}).
The bound on the size follows from the structure of spheres.
\qed
\end{proof}


The 
number of equivalences generated on line 16 of Algorithm~\ref{algo:terminating}
(and therefore also the number of transitions leaving $\psphere$) may be large.
Many of them are, however, infeasible (cannot be taken in any reachable configuration of $\aut^d$),
and could be removed.
In most cases, the majority of such infeasible transitions may be identified locally, 
taking advantage of the invariant of all reachable configurations of $\aut^d$ enforced by Algorithm~\ref{algo:terminating}: 
namely, values of distinct parameters are always pairwise distinct.
Therefore, before building a transition for an equivalence $\sim$, 
we ask whether the $\sim$-equivalent assignment terms may indeed be made
equivalent assuming that the constructed transition guard~$g_i$
and---importantly---also the distinctness invariant hold right before the transition is taken. 
Technically, we create new transitions only from those equivalences $\sim$ such that 
$\issatof{\bigwedge_{p_1,p_2\in P_\psphere, \distof{p_1, p_2}} (p_1\neq p_2)
\land g_i \land \varphi_\sim}$ where $P_\psphere$ is the set of parameters of
$\psphere$ and $\distof{p_1, p_2}$ holds iff $p_1$ and $p_2$ are distinct
parameters.


\vspace*{-2.0mm}
\subsection{Reachability-Restricted CA Determinisation}\label{sec:reachRestr}

Above, we have described a terminating algorithm for CA determinisation.
While it is witnessed by our experiments 
that the algorithm
often generates much smaller automata than what could be obtained by transforming
the automata into NFAs and determinising them, a natural question is whether the
generated DCA is \emph{always} smaller or equal in size to the DFA built by
getting rid of the counters and using classical determinisation.
Unfortunately, the answer to this question is no.
The reason is that the transformation to a~DCA needs not recognise that some
generated transitions can never be executed and that some spheres are not
reachable.
To see this, it is enough to imagine a transition setting some counter $c$ to
zero and the only successor transition testing whether $c$ is positive.
The latter transition would not be executed when generating the DFA due to
working with concrete values of counters, but it would be considered when
constructing the DCA (since the construction does not know the values of the
counters).

In our experiments with CAs obtained from real-life regexes, the
above was not a~problem, but we note that, for the price of an increased cost of
the construction, one could further improve the algorithm by taking into account
some reachability information.
%
In an extreme case, one could first generate the DFA corresponding to the given
CA and then use it when generating the DCA (as a hopefully more compact
representation of the DFA).
In particular, whenever adding some new sphere into the DCA being built, the
algorithm can check whether there is a subset of states in the original CA
represented as a state of the DFA that is an instance of the sphere.
If not, the sphere is not added.
The resulting DCA can then never be bigger than the DFA since each control
state of the DFA (i.e., a subset of states of the original CA) is represented
by a single sphere only, likewise each transition of the DFA is represented by a
single transition of the DCA, and there are not any unreachable spheres or
transitions that cannot be executed.

Notice 
that the reachability pruning is an alternative to Algorithm~\ref{algo:terminating}. 
Algorithm~\ref{algo:basic} equipped with the reachability analysis is guaranteed to terminate. 
For example, when run on the CA in Fig.~\ref{fig:littleA}, it would generate a DCA isomorphic to that from Fig.~\ref{fig:littleAdet}.

\section{Monadic Counting}\label{sec:monadic}
\newcommand{\trans}[3]{\movestd{#1}{#2}{#3}}
\renewcommand{\secref}[1]{Section~\ref{#1}}


We now provide a simplified and more efficient version of the determinisation
algorithm.
The simplified version targets CAs that naturally arise from \emph{monadic
regexes}, i.e., regular expressions extended with counting limited to
\emph{character classes}.
Their abstract syntax is\vspace*{-1mm}
\begin{equation*}
R ::= ~\emptyset \mid \emp \mid \sguard \mid R_1R_2 \mid R_1 + R_2 \mid
R{*} \mid \sguard\{n,m\}\\[-1mm]
\end{equation*}
where $\sguard$ is a predicate denoting a set of alphabet
symbols, i.e., a~\emph{character class} ($\sguard$~will be used to denote character classes from now on), and $n,m\geq 0$ are integers.
The semantics is defined as usual, with $\sguard\{n,m\}$ denoting a string $w$
with $n\leq |w| \leq m$ symbols satisfying~$\sguard$.

The specialised determinisation algorithm is of a high practical relevance since
the monadic class is very common, as witnessed by our experiments, where it
covers over 95\,\% of the regexes with counting that we found (cf.~\secref{sec:experiments}).

\vspace{-2.0mm}
\subsection{Monadic Counting Automata}

Monadic regexes can be easily compiled to nondeterministic monadic CAs
satisfying certain structural properties summarised below.\footnote{We note that
we restrict ourselves to range sub-expressions of the form $\sguard\{n,n\}$ or
$\sguard\{0,n\}$ only.
This is without loss of generality since a general range expression
$\sguard\{m,n\}$ can be rewritten as $\sguard\{m,m\}.\sguard\{0,n-m\}$.}
In particular, a (nondeterministic) \emph{monadic counting automaton (MCA)} is
a~CA $\aut = (Q,C,I,F,\trel)$ where~the~following~holds:

1. The set $Q$ of control states is partitioned into a set of \emph{simple
states} $Q_s$ and a set of \emph{counting states} $Q_c$, i.e.,
$Q = Q_s \uplus Q_c$.

2. The set of counters~$C = \{c_q\mid q\in Q_c\}$ consists of a~unique counter
$c_q$ for every counting state~$q \in Q_c$.

3. All transitions containing counter guards or updates must be incident with
a~counting state in the following manner.
Every counting state $q \in Q_c$ has a~single \emph{increment transition},
a~self-loop $\trans{q}{\sguard \land c_q<\max_q,c_q'=c_q+1}{q}$ with the value of
$c_q$ limited by the \emph{bound} $\max_q$ of $q$, and possibly several
\emph{entry transitions} of the form $\trans{r}{\bar\sguard \land c_q'=0}{q}$, which
set $c_q$ to~$0$.
As for \emph{exit transitions}, every counting state is either \emph{exact} or
\emph{range}, where exact counting states have exit transitions of the form
$\trans{q}{\sguard \land c_q = \max_q}{s}$, and \emph{range} counting states have exit
transitions of the form 
$\trans{q}{\sguard,\top}{s}$ with $s\in Q$ s.t.~$s \neq q$.
That is, an exact counting state may be left only after exactly $\max_q$
repetitions of the incrementing transition (it corresponds to a~regular
expression \texttt{$\sguard$\{k\}}), while a range counting state may be left
sooner (it corresponds to a regular expression \texttt{$\sguard$\{0,k\}}).
We denote the set of range counting states $Q_r$ and the set of exact counting
states $Q_e$,
with $Q_c = Q_r \uplus Q_e$.

4. The initial condition $I$ is of the form $I:\bigvee_{q\in Q_s^I} \stvar=q \lor
\bigvee_{q\in Q_c^I} (\stvar =q \land c_q = 0)$ for some sets of initial simple
and counting states $Q_s^I \subseteq Q_s$ and $Q_c^I \subseteq Q_c$,
respectively, with the counters of initial counting states initialised to $0$.

5. The final condition $F$ is of the form $F:\bigvee_{q\in Q_s^F \cup Q_r^F}
\stvar = q \lor \bigvee_{q\in Q_e^F} (\stvar =q \land c_q = \max_q)$ where
$Q_s^F \subseteq Q_s$ is a set of simple final states, $Q_r^F \subseteq Q_r$ is
a set of final range counting states, and $Q_e^F \subseteq Q_e$ is a set of
final exact counting states.
That is, final conditions on final states are the same as counter conditions on
exit transitions.\footnote{Notice that the guards $c_q < \max_q$ on the
incrementing self-loops of exact counting states could be removed without affecting
the language since when $c_q$ exceeds $\max_q$, then the run can never leave
$q$ and has thus no chance of accepting.
We include these guards only to conform to the condition on boundedness of counter
values in the definition of CAs.}

\subsection{Determinisation of MCAs}
\label{sec:monadicalgo}

Algorithm~\ref{algo:terminating} can be simplified when specialised to monadic
CAs. 
The simplification is based on the following observations.
\emph{Observation 1. Counters are dead outside their states.}
To simplify the representation of spheres, we use the fact that every counter
$c_q$ of an MCA is ``active'' in the state $q$ only, while $c_q$ is ``dead'' in
other states (i.e., its current value has no influence on runs of the MCA that
are not in~$q$). 
To represent different variants of $c_q$, we use parameters of the form $c_q[i]$
obtained by indexing $c_q$ by an index $i$, for $0\leq i \leq \max_q$, while
enforcing the invariant that, for distinct indices $i$ and $j$, $c_q[i]$ and
$c_q[j]$ always have different values.
Since the value of $c_q$ ranges from $0$ to $\max_q$, at most $\max_q+1$
variants of $c_q$ are needed.%
\footnote{Notice that maintaining a fixed association of a parameter to a counter is a
difference from Algorithms~\ref{algo:basic} and \ref{algo:terminating}, where one
parameter may represent different counters.}
Since spheres only need parameters to remember values of live counters,
every sphere can be equivalently written in the \emph{normal form} 
\begin{equation*}
  \psphere \defequiv \bigvee_{q \in Q'_s} \stvar = q ~\lor
\bigvee_{q \in Q'_c} \Big(\stvar = q \land \bigvee_{0 \leq i \leq \max'_{q}} c_q =
c_q[i]\Big)
\end{equation*}
for some $Q'_s \subseteq Q_s$, $Q'_c \subseteq Q_c$, and $\max'_{q} \leq
\max_{q}$.
That is, a sphere $\psphere$ records which states may be reached in the original
MCA when $\psphere$ is reached in the determinised MCA and also which variants
of the counter $c_q$ may record the value of $c_q$ when $q$ is reached.

\emph{Observation 2. Variants of exact counting states can be sorted.}
For dealing with any exact counting state $q \in Q_e$, we may use the
following facts:
(1) If executed, the increment transition of~$q$ increments all variants of $c_q$
whose values are smaller than~$\max_q$.
(2)~New variants of~$c_q$ are initialised to~$0$ by the entry transitions.
(3)~Variants whose value is $\max_q$ can take an exit transition, after which
they become dead and their values do not need to be propagated to the next
configuration. 
It is therefore easy to enforce that the values of the variants $c_q[i]$ stay
sorted, so that $i<j$ implies $\conf(c_q[i])<\conf(c_q[j])$ in every
configuration $\conf$ of $A^d$.
The sortedness invariant implies that the variant of $c_q$ with the highest index, called \emph{highest variant}, has the highest value. 
This, together with the invariant of boundedness by $\max_q$ and mutual distinctness of values of variants of $c_q$, means that the highest variant is the only one that may satisfy the tests $c_q = \max_q$ on exit transitions or fail the test $c_q <\max_q$ on the incrementing transition.
Hence, the deterministic MCA does not need to test all variants of $c_q$ but
the highest one only.

\emph{Observation 3. Only the smallest variants of range counting states are important.}
For range counting states, we adapt the \emph{simulation pruning} technique
from~\cite{5algos}.
The technique optimizes the standard subset-construction-based determinisation
of NFAs by exploiting a~\emph{simulation}
relation~\cite{DillHH91} such that any \emph{macrostate} (which
has the form of a~set of states of the original NFA) obtained during the
determinisation can be pruned by removing those NFA states that are simulated by
other NFA states included in the same macrostate.
The pruning does not change the language: the resulting DFA is bisimilar to the
one constructed without pruning.
For our DCA construction, we use the simulation that implicitly exists between
configurations $\conf$ and $\conf'$ of $\aut$ with the same range counting state $q =
\conf(\stvar) = \conf'(\stvar)$, where $\conf(c_q)\geq \conf'(c_q)$ implies
that~$\conf'$ simulates~$\conf$.\footnote{The fact that this relation is indeed
a~simulation can be seen from that both the higher and lower value of $c_q$ can
use any exit transition of $q$ at any moment regardless of the value of $c_q$,
but the lower value of $c_q$ can stay in the counting loop longer.}
Hence, the spheres only need to remember the smallest possible counter value for
every range counting state $q$, which may be always stored in $c_q[0]$, 
and discard all other variants.


\paragraph{Determinisation of MCAs.}
Observations 1--3 above allow for representing spheres using a simple data
structure, namely, a multiset of states. 
By a~slight abuse of notation, we use~$\psphere$ for the sphere itself as well as for its multiset representation 
$\fsphere: Q\rightarrow \nat$. 
The fact that $\fsphere(q) > 0$ means that $q$ is present in the sphere
(i.e., $\stvar = q$ is a predicate in the normal form of $\psphere$),
and for a counting state~$q$, the counters
$c_q[0],\ldots,c_q[\fsphere(q)-1]$ are the $\fsphere(q)$ variants of
$c_q$ tracked in the sphere \mbox{(i.e.,
$\fsphere(q)-1 = \max_q'$ in the normal form of $\psphere$).}

The MCA determinisation is then an analogy of Algorithm~\ref{algo:basic} that
uses the multiset data structure and preserves the sortedness and uniqueness of
variants of exact counters.
The initial sphere $\fsphere_I$ assigns~$1$ to all initial states of~$I$, and
the initial configuration $I^d$ assigns~$0$ to~$c_q[0]$ for each counting
state~$q$ in~$I$.
Further, we modify the part of Algorithm~\ref{algo:basic} after popping
a~sphere~$\psphere$ from $\wl$ in the main loop (lines~\ref{ln:factorise}--\ref{ln:addQ}).

Let $\trel_\fsphere$ denote the set of transitions of $\aut$ originating from
states~$q$ with $\fsphere(q) > 0$.
Processing of~$\fsphere$ starts by removing guard predicates of the form $c_q <
\max_q$ from increment transitions of exact counting states in $\trel_\fsphere$ (since
they have no semantic effect as mentioned already above).
Subsequently, we compute minterms of the set of guard formulae of the transitions in $\trel_\fsphere$.
Each minterm $\minterm$ then gives rise to a transition
$\trans{\fsphere}{\guard,\assignments}{\fsphere'}$ of~$\aut^d$.
The guard formula~$\guard$, assignment formula~$\assignments$, 
and the target sphere~$\fsphere'$ are constructed as follows.

First, the guard~$\guard$ is obtained from the minterm~$\minterm$ by replacing,
for all $q \in Q_c$, every occurrence of $c_q$ by $c_q[\fsphere(q)]$, i.e.,
the highest variant of~$c_q$.
Intuitively, the counter guards of transitions of $\trel_\fsphere$ present in $\minterm$ will on the constructed transition of $\aut^d$ be testing the highest variants of the counters. 
This is justified since (a)~only the highest variant of~$c_q$ needs to be tested
for exact counting states, as concluded in Observation 2 above, and (b)~we keep only a~single variant of~$c_q$ for
range counting states (which is also the highest one), as concluded in Observation 3.

We then initialise the target multiset $\fsphere'$ as the empty multiset $\{q\mapsto 0\mid q\in Q\}$
and collect the set $\trel_\minterm$ of all transitions
from~$\trel_\fsphere$ that are compatible with the
minterm~$\minterm$ 
(recall that increment self-loops of exact states in $\trel_\fsphere$ have counter guards removed, 
hence counter guards do not influence their inclusion in $\trel_\minterm$).
The transitions of $\trel_\minterm$ will be processed in the following three steps. 

\emph{Step 1 (simple states).}
Simple states with an incoming transition in $\trel_\minterm$ get $\fsphere'(q) = 1$.

\emph{Step 2 (increment self-loops).} For exact states with the increment
self-loop in $\trel_\minterm$, $\fsphere'(q)$ is set to $\fsphere(q)-1$ if an
exit transition of $q$ is in $\trel_\minterm$, and to $\fsphere(q)$ otherwise.
Indeed, if (and only if) an exit transition of $q$ is included in
$\trel_\minterm$, and $\trel_\minterm$ is enabled in some sphere, then the
highest variant of $c_q$ has reached $\max_q$ in that sphere, and the self-loop
cannot be taken by the highest variant of $c_q$.
%
%
The lower variants of $c_q$ always have values smaller than $\max_q$, and hence
can take the self-loop. 
The assignment $\assignments$ then gets the~conjunct $c_q[i]' = c_q[i]+1$ for
each $0\leq i < \fsphere'(q)$ since the variants that take the self-loop are
incremented.
For range states with the increment self-loop in $\trel_\minterm$,
we set $\fsphere'(q)$ to 1, and $c_q[0]' = c_q[0]+1$ is added to $\assignments$
(only one variant is remembered).

\emph{Step 3 (entry transitions).}
For each counting state $q$ with an entry transition in $\trel_\minterm$, 
$\fsphere'(q)$ is incremented by $1$ and the assignment $c_q[0]' = 0$ of the fresh variant of $c_q$ is added to $\assignments$. 
If the new value of $\fsphere'(q)$ exceeds $\max_q+1$, then the whole transition generated from $\minterm$ is discarded, since $c_q$ cannot have more than $\max_q + 1$ distinct values.
Otherwise, if $q$ is an exact counting state, 
then $\assignments$ is updated to preserve the invariant of sorted and unique
values of $c_q$:
the increments of older variants of $c_q$ are
\emph{right-shifted} to make space for the fresh variant, 
meaning that each conjunct $c_q[i]' = c_q[i]+1$ in $f$ is replaced by $c_q[i+1]' = c_q[i]+1$. 
If $q \in Q_r$, 
then if the assignment $c_q[0]' = c_q[0]+1$ is present in $\assignments$, it is removed 
(as the fresh variant has the \mbox{smallest value $0$).}

\begin{example}
Determinising 
the CA from Fig.~\ref{fig:littleA} using the algorithm described in this section
would result in the DCA shown in
Fig.~\ref{fig:littleAdet}. 
%
\qed
\end{example}

The monadic determinisation has a much lower worst-case complexity than the general algorithm. 
Importantly, 
the number of states depends on $\max_A$ only polynomially,
which is a major difference from the exponential bounds of the na\"{i}ve determinisation and our general construction.

\begin{theorem}
  The specialised monadic CA determinisation constructs a~DCA with $|Q^d| \leq
  (\max_\aut + 1)^{|Q|}$ and $|\trel^d| \leq |\Sigma| \cdot (4\cdot
  (\max_\aut+1))^{|Q|}$.
\end{theorem}

\begin{proof}[idea] The bound on the number of states is given by the number of
functions $Q \to \{0, \ldots,  \max_\aut\}$.
The bound on the number of transitions is given by the fact, that if a~sphere
multiset maps a~state $q$ to $n$, then the successors of the sphere can map $q$
to~$0$ (when $q$ is not a successor), $n-1$, $n$, or $n+1$.
Therefore, for every symbol from~$\Sigma$ and every macrostate from at most
$(\max_\aut+1)^{|Q|}$ many of them, there are at most $4^{|Q|}$ successors, and
$|\Sigma| \cdot (\max_\aut+1)^{|Q|} \cdot 4^{|Q|} = (4\cdot
(\max_\aut+1))^{|Q|}$.
\qed
\end{proof}


\vspace{-3.0mm}
\section{Experimental Evaluation}\label{sec:experiments}
\vspace{-2.0mm}

The main purpose of our experimentation was to compare the proposed approach
with the na\"{i}ve determinisation and confirm that our method produces
significantly smaller automata and mitigates the risk of the state space
explosion causing a complete failure of determinisation 
(and the implied
impossibility to use the desired deterministic automaton for the intended
application, such as pattern matching).
To this end, we extended the Microsoft's Automata
library~\cite{automata_library} with a~prototype support for CAs, implemented
the algorithm from \secref{sec:monadic} (denoted \textbf{Counting} in the
following), and compared it to the standard determinisation already present in
the library (denoted as \textbf{DFA}).
For the evaluation, we collected 2,361~regexes from a~wide range of
applications---namely, those used in network intrusion detection systems
(Snort~\cite{snort}: 741~regexes, Yang~\cite{yang2010}: 228~regexes,
Bro~\cite{bro}: 417~regexes, HomeBrewed~\cite{tacas18-appred}: 55~regexes), the
Microsoft's security leak scanning system~(Industrial: 17~regexes), the Sagan
log analysis engine (Sagan~\cite{sagan}: 14~regexes), and the pattern matching
rules from RegExLib (RegExLib~\cite{regexlib}: 889~regexes).
%
We only selected regexes that contain an occurrence of the counting operator,
and from these, we selected only monadic ones (there were over~95\,\% of them,
confirming the fragment's importance). All benchmarks were run on a Xeon
E5-2620v2@2.4GHz CPU with 32\,GiB RAM with a~timeout of 1~min (we take
the~mean time of 10~runs).
In the following, we use $\mean$, $\median$, and $\stdev$ to denote the
statistical indicators mean, median, and standard deviation, respectively.
All times are reported in~milliseconds.

\begin{wrapfigure}[14]{r}{4.7cm}
\vspace*{-9mm}
\hspace*{-3mm}
\begin{minipage}{5cm}
\input{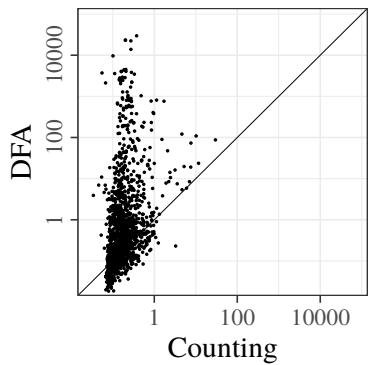}
\end{minipage}
\vspace{-10mm}
\caption{Comparison of running times given in ms (the axes are
  logarithmic).}
\label{fig:runtimes}
\end{wrapfigure}
The number of timeouts was 
110 for \textbf{Counting}, and 238 for \textbf{DFA}.  The two methods were to
some degree complementary, there were only 62 cases in which both timed out.
This confirms that our
algorithm indeed mitigates the risk of failure due to state space explosion in
determinisation.  The remaining comparisons are done only with respect to
benchmarks for which \mbox{neither of the methods timed out.}


In Fig.~\ref{fig:runtimes}, we compare the running times of the conversion of an
NFA for a~given regex to a~DFA (the \textbf{DFA} axis) and the determinisation
of the CA for the same regex (the \textbf{Counting} axis). 
If we exclude the easy cases where both approaches finished within 1\,ms, 
we can see that \textbf{Counting} is almost
always better than \textbf{DFA}.
Note that the axes are logarithmic, so the advantage of \textbf{Counting} over
\textbf{DFA} grows exponentially wrt the distance of the data point from the
diagonal. 
The statistical indicators for the running times are 
$\mean=110$, $\median=0.17$, $\stdev=1,177$ 
for \textbf{DFA} and
$\mean=0.23$, $\median=0.13$, $\stdev=0.09$ for \textbf{Counting}.

\begin{wrapfigure}[14]{r}{4.7cm}
\vspace*{-9mm}
\hspace*{-3mm}
\begin{minipage}{5cm}
\input{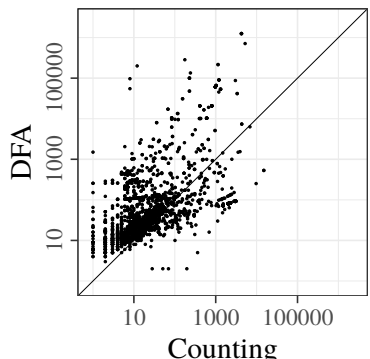}
\end{minipage}
\vspace{-10mm}
\caption{Comparison of numbers of states (the axes are logarithmic).}
\label{fig:states}
\end{wrapfigure}
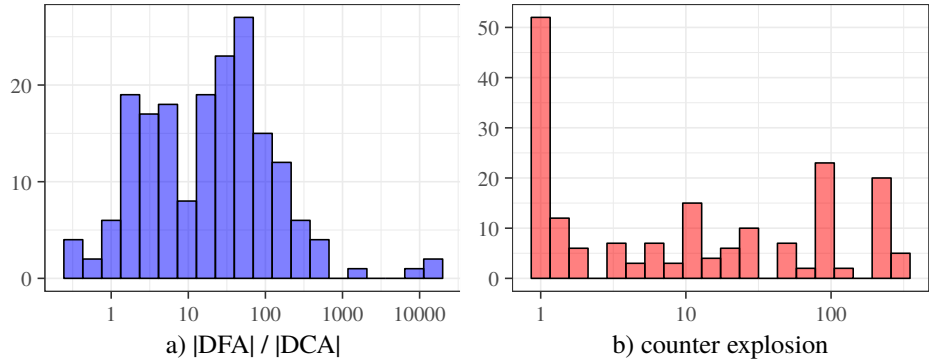
\begin{figure}[t]
  \begin{minipage}[b]{0.5\linewidth}
  \begin{center}
\begin{tikzpicture}[x=1pt,y=1pt]
\definecolor{fillColor}{RGB}{255,255,255}
\path[use as bounding box,fill=fillColor,fill opacity=0.00] (0,0) rectangle (180.67,144.54);
\begin{scope}
\path[clip] (  0.00,  0.00) rectangle (180.67,144.54);
\definecolor{drawColor}{RGB}{255,255,255}
\definecolor{fillColor}{RGB}{255,255,255}

\path[draw=drawColor,line width= 0.6pt,line join=round,line cap=round,fill=fillColor] (  0.00, -0.00) rectangle (180.68,144.54);
\end{scope}
\begin{scope}
\path[clip] ( 19.25, 30.72) rectangle (175.17,139.04);
\definecolor{fillColor}{RGB}{255,255,255}

\path[fill=fillColor] ( 19.25, 30.72) rectangle (175.17,139.04);
\definecolor{drawColor}{gray}{0.92}

\path[draw=drawColor,line width= 0.3pt,line join=round] ( 19.25, 53.88) --
	(175.17, 53.88);

\path[draw=drawColor,line width= 0.3pt,line join=round] ( 19.25, 90.35) --
	(175.17, 90.35);

\path[draw=drawColor,line width= 0.3pt,line join=round] ( 19.25,126.82) --
	(175.17,126.82);

\path[draw=drawColor,line width= 0.3pt,line join=round] ( 29.63, 30.72) --
	( 29.63,139.04);

\path[draw=drawColor,line width= 0.3pt,line join=round] ( 58.48, 30.72) --
	( 58.48,139.04);

\path[draw=drawColor,line width= 0.3pt,line join=round] ( 87.33, 30.72) --
	( 87.33,139.04);

\path[draw=drawColor,line width= 0.3pt,line join=round] (116.18, 30.72) --
	(116.18,139.04);

\path[draw=drawColor,line width= 0.3pt,line join=round] (145.04, 30.72) --
	(145.04,139.04);

\path[draw=drawColor,line width= 0.3pt,line join=round] (173.89, 30.72) --
	(173.89,139.04);

\path[draw=drawColor,line width= 0.6pt,line join=round] ( 19.25, 35.65) --
	(175.17, 35.65);

\path[draw=drawColor,line width= 0.6pt,line join=round] ( 19.25, 72.12) --
	(175.17, 72.12);

\path[draw=drawColor,line width= 0.6pt,line join=round] ( 19.25,108.59) --
	(175.17,108.59);

\path[draw=drawColor,line width= 0.6pt,line join=round] ( 44.05, 30.72) --
	( 44.05,139.04);

\path[draw=drawColor,line width= 0.6pt,line join=round] ( 72.91, 30.72) --
	( 72.91,139.04);

\path[draw=drawColor,line width= 0.6pt,line join=round] (101.76, 30.72) --
	(101.76,139.04);

\path[draw=drawColor,line width= 0.6pt,line join=round] (130.61, 30.72) --
	(130.61,139.04);

\path[draw=drawColor,line width= 0.6pt,line join=round] (159.46, 30.72) --
	(159.46,139.04);
\definecolor{drawColor}{RGB}{0,0,0}
\definecolor{fillColor}{RGB}{0,0,255}

\path[draw=drawColor,line width= 0.6pt,line join=round,fill=fillColor,fill opacity=0.50] ( 26.34, 35.65) rectangle ( 33.42, 50.24);

\path[draw=drawColor,line width= 0.6pt,line join=round,fill=fillColor,fill opacity=0.50] ( 33.42, 35.65) rectangle ( 40.51, 42.94);

\path[draw=drawColor,line width= 0.6pt,line join=round,fill=fillColor,fill opacity=0.50] ( 40.51, 35.65) rectangle ( 47.60, 57.53);

\path[draw=drawColor,line width= 0.6pt,line join=round,fill=fillColor,fill opacity=0.50] ( 47.60, 35.65) rectangle ( 54.69,104.94);

\path[draw=drawColor,line width= 0.6pt,line join=round,fill=fillColor,fill opacity=0.50] ( 54.69, 35.65) rectangle ( 61.77, 97.65);

\path[draw=drawColor,line width= 0.6pt,line join=round,fill=fillColor,fill opacity=0.50] ( 61.77, 35.65) rectangle ( 68.86,101.29);

\path[draw=drawColor,line width= 0.6pt,line join=round,fill=fillColor,fill opacity=0.50] ( 68.86, 35.65) rectangle ( 75.95, 64.82);

\path[draw=drawColor,line width= 0.6pt,line join=round,fill=fillColor,fill opacity=0.50] ( 75.95, 35.65) rectangle ( 83.04,104.94);

\path[draw=drawColor,line width= 0.6pt,line join=round,fill=fillColor,fill opacity=0.50] ( 83.04, 35.65) rectangle ( 90.12,119.53);

\path[draw=drawColor,line width= 0.6pt,line join=round,fill=fillColor,fill opacity=0.50] ( 90.12, 35.65) rectangle ( 97.21,134.12);

\path[draw=drawColor,line width= 0.6pt,line join=round,fill=fillColor,fill opacity=0.50] ( 97.21, 35.65) rectangle (104.30, 90.35);

\path[draw=drawColor,line width= 0.6pt,line join=round,fill=fillColor,fill opacity=0.50] (104.30, 35.65) rectangle (111.39, 79.41);

\path[draw=drawColor,line width= 0.6pt,line join=round,fill=fillColor,fill opacity=0.50] (111.39, 35.65) rectangle (118.47, 57.53);

\path[draw=drawColor,line width= 0.6pt,line join=round,fill=fillColor,fill opacity=0.50] (118.47, 35.65) rectangle (125.56, 50.24);

\path[draw=drawColor,line width= 0.6pt,line join=round,fill=fillColor,fill opacity=0.50] (125.56, 35.65) rectangle (132.65, 35.65);

\path[draw=drawColor,line width= 0.6pt,line join=round,fill=fillColor,fill opacity=0.50] (132.65, 35.65) rectangle (139.74, 39.30);

\path[draw=drawColor,line width= 0.6pt,line join=round,fill=fillColor,fill opacity=0.50] (139.74, 35.65) rectangle (146.82, 35.65);

\path[draw=drawColor,line width= 0.6pt,line join=round,fill=fillColor,fill opacity=0.50] (146.82, 35.65) rectangle (153.91, 35.65);

\path[draw=drawColor,line width= 0.6pt,line join=round,fill=fillColor,fill opacity=0.50] (153.91, 35.65) rectangle (161.00, 39.30);

\path[draw=drawColor,line width= 0.6pt,line join=round,fill=fillColor,fill opacity=0.50] (161.00, 35.65) rectangle (168.09, 42.94);
\definecolor{drawColor}{gray}{0.20}

\path[draw=drawColor,line width= 0.6pt,line join=round,line cap=round] ( 19.25, 30.72) rectangle (175.17,139.04);
\end{scope}
\begin{scope}
\path[clip] (  0.00,  0.00) rectangle (180.67,144.54);
\definecolor{drawColor}{gray}{0.30}

\node[text=drawColor,anchor=base east,inner sep=0pt, outer sep=0pt, scale=  0.88] at ( 14.30, 32.62) {0};

\node[text=drawColor,anchor=base east,inner sep=0pt, outer sep=0pt, scale=  0.88] at ( 14.30, 69.09) {10};

\node[text=drawColor,anchor=base east,inner sep=0pt, outer sep=0pt, scale=  0.88] at ( 14.30,105.56) {20};
\end{scope}
\begin{scope}
\path[clip] (  0.00,  0.00) rectangle (180.67,144.54);
\definecolor{drawColor}{gray}{0.20}

\path[draw=drawColor,line width= 0.6pt,line join=round] ( 16.50, 35.65) --
	( 19.25, 35.65);

\path[draw=drawColor,line width= 0.6pt,line join=round] ( 16.50, 72.12) --
	( 19.25, 72.12);

\path[draw=drawColor,line width= 0.6pt,line join=round] ( 16.50,108.59) --
	( 19.25,108.59);
\end{scope}
\begin{scope}
\path[clip] (  0.00,  0.00) rectangle (180.67,144.54);
\definecolor{drawColor}{gray}{0.20}

\path[draw=drawColor,line width= 0.6pt,line join=round] ( 44.05, 27.97) --
	( 44.05, 30.72);

\path[draw=drawColor,line width= 0.6pt,line join=round] ( 72.91, 27.97) --
	( 72.91, 30.72);

\path[draw=drawColor,line width= 0.6pt,line join=round] (101.76, 27.97) --
	(101.76, 30.72);

\path[draw=drawColor,line width= 0.6pt,line join=round] (130.61, 27.97) --
	(130.61, 30.72);

\path[draw=drawColor,line width= 0.6pt,line join=round] (159.46, 27.97) --
	(159.46, 30.72);
\end{scope}
\begin{scope}
\path[clip] (  0.00,  0.00) rectangle (180.67,144.54);
\definecolor{drawColor}{gray}{0.30}

\node[text=drawColor,anchor=base,inner sep=0pt, outer sep=0pt, scale=  0.88] at ( 44.05, 19.71) {1};

\node[text=drawColor,anchor=base,inner sep=0pt, outer sep=0pt, scale=  0.88] at ( 72.91, 19.71) {10};

\node[text=drawColor,anchor=base,inner sep=0pt, outer sep=0pt, scale=  0.88] at (101.76, 19.71) {100};

\node[text=drawColor,anchor=base,inner sep=0pt, outer sep=0pt, scale=  0.88] at (130.61, 19.71) {1000};

\node[text=drawColor,anchor=base,inner sep=0pt, outer sep=0pt, scale=  0.88] at (159.46, 19.71) {10000};
\end{scope}
\begin{scope}
\path[clip] (  0.00,  0.00) rectangle (180.67,144.54);
\definecolor{drawColor}{RGB}{0,0,0}

\node[text=drawColor,anchor=base,inner sep=0pt, outer sep=0pt, scale=  1.10] at ( 97.21,  7.44) {a) |DFA| / |DCA|};
\end{scope}
\end{tikzpicture}
  \end{center}
  \end{minipage}
  \begin{minipage}[b]{0.5\linewidth}
  \begin{center}
\begin{tikzpicture}[x=1pt,y=1pt]
\definecolor{fillColor}{RGB}{255,255,255}
\path[use as bounding box,fill=fillColor,fill opacity=0.00] (0,0) rectangle (180.67,144.54);
\begin{scope}
\path[clip] (  0.00,  0.00) rectangle (180.67,144.54);
\definecolor{drawColor}{RGB}{255,255,255}
\definecolor{fillColor}{RGB}{255,255,255}

\path[draw=drawColor,line width= 0.6pt,line join=round,line cap=round,fill=fillColor] (  0.00, -0.00) rectangle (180.68,144.54);
\end{scope}
\begin{scope}
\path[clip] ( 19.25, 30.72) rectangle (175.17,139.04);
\definecolor{fillColor}{RGB}{255,255,255}

\path[fill=fillColor] ( 19.25, 30.72) rectangle (175.17,139.04);
\definecolor{drawColor}{gray}{0.92}

\path[draw=drawColor,line width= 0.3pt,line join=round] ( 19.25, 45.12) --
	(175.17, 45.12);

\path[draw=drawColor,line width= 0.3pt,line join=round] ( 19.25, 64.05) --
	(175.17, 64.05);

\path[draw=drawColor,line width= 0.3pt,line join=round] ( 19.25, 82.99) --
	(175.17, 82.99);

\path[draw=drawColor,line width= 0.3pt,line join=round] ( 19.25,101.92) --
	(175.17,101.92);

\path[draw=drawColor,line width= 0.3pt,line join=round] ( 19.25,120.86) --
	(175.17,120.86);

\path[draw=drawColor,line width= 0.3pt,line join=round] ( 57.05, 30.72) --
	( 57.05,139.04);

\path[draw=drawColor,line width= 0.3pt,line join=round] (111.38, 30.72) --
	(111.38,139.04);

\path[draw=drawColor,line width= 0.3pt,line join=round] (165.71, 30.72) --
	(165.71,139.04);

\path[draw=drawColor,line width= 0.6pt,line join=round] ( 19.25, 35.65) --
	(175.17, 35.65);

\path[draw=drawColor,line width= 0.6pt,line join=round] ( 19.25, 54.58) --
	(175.17, 54.58);

\path[draw=drawColor,line width= 0.6pt,line join=round] ( 19.25, 73.52) --
	(175.17, 73.52);

\path[draw=drawColor,line width= 0.6pt,line join=round] ( 19.25, 92.46) --
	(175.17, 92.46);

\path[draw=drawColor,line width= 0.6pt,line join=round] ( 19.25,111.39) --
	(175.17,111.39);

\path[draw=drawColor,line width= 0.6pt,line join=round] ( 19.25,130.33) --
	(175.17,130.33);

\path[draw=drawColor,line width= 0.6pt,line join=round] ( 29.88, 30.72) --
	( 29.88,139.04);

\path[draw=drawColor,line width= 0.6pt,line join=round] ( 84.21, 30.72) --
	( 84.21,139.04);

\path[draw=drawColor,line width= 0.6pt,line join=round] (138.54, 30.72) --
	(138.54,139.04);
\definecolor{drawColor}{RGB}{0,0,0}
\definecolor{fillColor}{RGB}{255,0,0}

\path[draw=drawColor,line width= 0.6pt,line join=round,fill=fillColor,fill opacity=0.50] ( 26.34, 35.65) rectangle ( 33.42,134.12);

\path[draw=drawColor,line width= 0.6pt,line join=round,fill=fillColor,fill opacity=0.50] ( 33.42, 35.65) rectangle ( 40.51, 58.37);

\path[draw=drawColor,line width= 0.6pt,line join=round,fill=fillColor,fill opacity=0.50] ( 40.51, 35.65) rectangle ( 47.60, 47.01);

\path[draw=drawColor,line width= 0.6pt,line join=round,fill=fillColor,fill opacity=0.50] ( 47.60, 35.65) rectangle ( 54.69, 35.65);

\path[draw=drawColor,line width= 0.6pt,line join=round,fill=fillColor,fill opacity=0.50] ( 54.69, 35.65) rectangle ( 61.77, 48.90);

\path[draw=drawColor,line width= 0.6pt,line join=round,fill=fillColor,fill opacity=0.50] ( 61.77, 35.65) rectangle ( 68.86, 41.33);

\path[draw=drawColor,line width= 0.6pt,line join=round,fill=fillColor,fill opacity=0.50] ( 68.86, 35.65) rectangle ( 75.95, 48.90);

\path[draw=drawColor,line width= 0.6pt,line join=round,fill=fillColor,fill opacity=0.50] ( 75.95, 35.65) rectangle ( 83.04, 41.33);

\path[draw=drawColor,line width= 0.6pt,line join=round,fill=fillColor,fill opacity=0.50] ( 83.04, 35.65) rectangle ( 90.12, 64.05);

\path[draw=drawColor,line width= 0.6pt,line join=round,fill=fillColor,fill opacity=0.50] ( 90.12, 35.65) rectangle ( 97.21, 43.22);

\path[draw=drawColor,line width= 0.6pt,line join=round,fill=fillColor,fill opacity=0.50] ( 97.21, 35.65) rectangle (104.30, 47.01);

\path[draw=drawColor,line width= 0.6pt,line join=round,fill=fillColor,fill opacity=0.50] (104.30, 35.65) rectangle (111.39, 54.58);

\path[draw=drawColor,line width= 0.6pt,line join=round,fill=fillColor,fill opacity=0.50] (111.39, 35.65) rectangle (118.47, 35.65);

\path[draw=drawColor,line width= 0.6pt,line join=round,fill=fillColor,fill opacity=0.50] (118.47, 35.65) rectangle (125.56, 48.90);

\path[draw=drawColor,line width= 0.6pt,line join=round,fill=fillColor,fill opacity=0.50] (125.56, 35.65) rectangle (132.65, 39.44);

\path[draw=drawColor,line width= 0.6pt,line join=round,fill=fillColor,fill opacity=0.50] (132.65, 35.65) rectangle (139.74, 79.20);

\path[draw=drawColor,line width= 0.6pt,line join=round,fill=fillColor,fill opacity=0.50] (139.74, 35.65) rectangle (146.82, 39.44);

\path[draw=drawColor,line width= 0.6pt,line join=round,fill=fillColor,fill opacity=0.50] (146.82, 35.65) rectangle (153.91, 35.65);

\path[draw=drawColor,line width= 0.6pt,line join=round,fill=fillColor,fill opacity=0.50] (153.91, 35.65) rectangle (161.00, 73.52);

\path[draw=drawColor,line width= 0.6pt,line join=round,fill=fillColor,fill opacity=0.50] (161.00, 35.65) rectangle (168.09, 45.12);
\definecolor{drawColor}{gray}{0.20}

\path[draw=drawColor,line width= 0.6pt,line join=round,line cap=round] ( 19.25, 30.72) rectangle (175.17,139.04);
\end{scope}
\begin{scope}
\path[clip] (  0.00,  0.00) rectangle (180.67,144.54);
\definecolor{drawColor}{gray}{0.30}

\node[text=drawColor,anchor=base east,inner sep=0pt, outer sep=0pt, scale=  0.88] at ( 14.30, 32.62) {0};

\node[text=drawColor,anchor=base east,inner sep=0pt, outer sep=0pt, scale=  0.88] at ( 14.30, 51.55) {10};

\node[text=drawColor,anchor=base east,inner sep=0pt, outer sep=0pt, scale=  0.88] at ( 14.30, 70.49) {20};

\node[text=drawColor,anchor=base east,inner sep=0pt, outer sep=0pt, scale=  0.88] at ( 14.30, 89.43) {30};

\node[text=drawColor,anchor=base east,inner sep=0pt, outer sep=0pt, scale=  0.88] at ( 14.30,108.36) {40};

\node[text=drawColor,anchor=base east,inner sep=0pt, outer sep=0pt, scale=  0.88] at ( 14.30,127.30) {50};
\end{scope}
\begin{scope}
\path[clip] (  0.00,  0.00) rectangle (180.67,144.54);
\definecolor{drawColor}{gray}{0.20}

\path[draw=drawColor,line width= 0.6pt,line join=round] ( 16.50, 35.65) --
	( 19.25, 35.65);

\path[draw=drawColor,line width= 0.6pt,line join=round] ( 16.50, 54.58) --
	( 19.25, 54.58);

\path[draw=drawColor,line width= 0.6pt,line join=round] ( 16.50, 73.52) --
	( 19.25, 73.52);

\path[draw=drawColor,line width= 0.6pt,line join=round] ( 16.50, 92.46) --
	( 19.25, 92.46);

\path[draw=drawColor,line width= 0.6pt,line join=round] ( 16.50,111.39) --
	( 19.25,111.39);

\path[draw=drawColor,line width= 0.6pt,line join=round] ( 16.50,130.33) --
	( 19.25,130.33);
\end{scope}
\begin{scope}
\path[clip] (  0.00,  0.00) rectangle (180.67,144.54);
\definecolor{drawColor}{gray}{0.20}

\path[draw=drawColor,line width= 0.6pt,line join=round] ( 29.88, 27.97) --
	( 29.88, 30.72);

\path[draw=drawColor,line width= 0.6pt,line join=round] ( 84.21, 27.97) --
	( 84.21, 30.72);

\path[draw=drawColor,line width= 0.6pt,line join=round] (138.54, 27.97) --
	(138.54, 30.72);
\end{scope}
\begin{scope}
\path[clip] (  0.00,  0.00) rectangle (180.67,144.54);
\definecolor{drawColor}{gray}{0.30}

\node[text=drawColor,anchor=base,inner sep=0pt, outer sep=0pt, scale=  0.88] at ( 29.88, 19.71) {1};

\node[text=drawColor,anchor=base,inner sep=0pt, outer sep=0pt, scale=  0.88] at ( 84.21, 19.71) {10};

\node[text=drawColor,anchor=base,inner sep=0pt, outer sep=0pt, scale=  0.88] at (138.54, 19.71) {100};
\end{scope}
\begin{scope}
\path[clip] (  0.00,  0.00) rectangle (180.67,144.54);
\definecolor{drawColor}{RGB}{0,0,0}

\node[text=drawColor,anchor=base,inner sep=0pt, outer sep=0pt, scale=  1.10] at ( 97.21,  7.44) {b) counter explosion};
\end{scope}
\end{tikzpicture}
  \end{center}
  \end{minipage}
\begin{center}
\end{center}
\vspace*{-14mm}
\caption{Histograms of (a) the ratio of the number of states of a~DFA and of the
  corresponding DCA (i.e., a~bar at value~$x$ of a height~$h$ denotes that
  the size of the DCA was $h$~times around $x$~times smaller than the size of the
  corresponding DFA) and
  (b) the ratio of the number of counters used by a~CA after and before
  determinisation.
  Note that the $x$-axes are logarithmic in both cases.}
  \vspace*{-4mm}
\label{fig:expl_histo}
\end{figure}

In Fig.~\ref{fig:states}, we compare the number of states of the results of the
determinisation algorithms (DCA for \textbf{Counting} and DFA for \textbf{DFA}).
Also here, \textbf{Counting} significantly dominates \textbf{DFA}.
The statistical indicators for the numbers of states are 
$\mean=4,543$, $\median=41$, $\stdev=57,543$
for \textbf{DFA} and
$\mean=241$, $\median=13$, $\stdev=800$ for \textbf{Counting}.
To better evaluate the conciseness of using DCAs,
we further selected 184 benchmarks that suffered from state explosion
during determinisation (our criterion for the selection was that the number of
states increased at least ten-fold in \textbf{DFA}) and explored how the CA
model can be used to mitigate the explosion.
Fig.~\ref{fig:expl_histo} shows histograms of how DCAs were more compact than
DFAs and also how much the number of counters rose during the determinisation.
From the histograms, we can see that there are indeed many cases where the
use of DCAs allows one to use a~significantly more compact representation, in some
cases by the factor of hundreds, thousands, or even tens of thousands.
Furthermore, the other histogram shows that, in many cases, no blow-up in the number
of counters happened; 
though there are also cases where
the number of counters increased by the factor of hundreds.

\begin{wrapfigure}[14]{r}{4.7cm}
\vspace*{-8mm}
\hspace*{-3mm}
\begin{minipage}{5cm}
\input{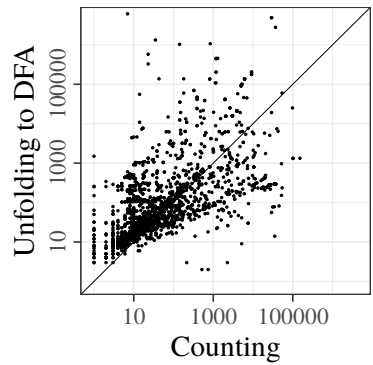}
\end{minipage}
\vspace{-10mm}
\caption{Comparison of numbers of transitions (the axes are logarithmic).}
\label{fig:trans}
\end{wrapfigure}

In terms of numbers of transitions, the methods compare similarly as for numbers of states,
as shown in Fig.~\ref{fig:trans}. 
We obtained
$\mean =14,282$,
$\median = 77$,
$\stdev =213,406$ for \textbf{DFA}
and
$\mean =2,398$,
$\median =23$,
$\stdev =8,475$ for \textbf{Counting}.
(We emphasize the number of states over the number of transitions in our comparisons since the performance and complexity of automata algorithms is usually more sensitive to the number of states, 
and large numbers of transitions are amenable for efficient symbolic representations \cite{KlaEtAl:Mona,SfaMinPOPL14,vata}.)

Benefits of the \textbf{Counting} method were the most substantial on the
Industrial dataset.
For the regex
"\verb=.*A[^AB]{0,800}C[D-G]{43,53}DFG[^D-H]="\\
(which was obtained from the real one, which is confidential, by
substituting the used character classes by characters \verb=A=--\verb=H=),
the obtained DFA contains 200,132 states, while the DCA contains only 12~states
(and 2~counters), which is 16,667 times less.
When minimised, the DFA still has 65,193 states.
There were other regexes where \textbf{Counting} achieved a~great reduction,
in total two regexes had a~reduction of over 10,000,
three more regexes had a~reduction of over 1,000, and 45~more had a~reduction of
over~100.

Additionally, we also compared our approach against the na\"{i}ve
determinisation followed by the standard minimisation. Due to the space
restrictions and since minimisation is not relevant to our primary target
(preventing failure due to state space explosion during determinisation), we
present the results only briefly.
Minimisation increased the running times of \textbf{DFA} by about one half
($\mean = 150$, $\median=0.35$, $\stdev = 1,582$ for the running times of
\textbf{DFA} followed by minimisation).  The minimal DFAs were on average about
ten times smaller than the original DFAs, and about ten times larger than our
DCAs ($\mean = 385$, $\median= 29$, $\stdev=4,195$ for the numbers of
states of the minimal DFAs).  

\vspace{-2.0mm}
\section{Related Work}\label{sec:related}
\vspace{-2.0mm}


Our notion of CAs is close to the definition of FACs in~\cite{Hovland09}, but our
CAs are more general, by allowing input predicates and more complex counter
updates.
Also $R$-automata~\cite{abdulla_rautomata} are related but somewhat orthogonal
to CAs because counters in $R$-automata do not need to have upper bounds and
cannot be tested or compared.
Counter systems are also related to CAs but allow more general operations over
counters through Presburger formulas~\cite{finkel_fast}.
CAs can also be seen as a special case of extended finite state machines or
EFSMs~\cite{cheng_efsm,shiple_efsm,jha_extended,Sperberg-McQueen-ExtendedFAWeb},
but these already go beyond regular languages.

Extended FAs (XFAs) augment classical automata with so-called scratch memory of
bits and bit-instructions~\cite{SEJ08,jha_extended}, which can represent counters
and also reduce nondeterminism.
%
%
Regexes are compiled into deterministic XFAs by first using an extended version
of Thompson's algorithm~\cite{Thom68}, then determinised through an extended version of the
classical powerset construction, and finally minimised.
Although a small XFA may exist, the determinisation algorithm incurs an
intermediate exponential blowup of search space for inputs such as
\texttt{.*a.\{$k$\}} (cf.~\cite[Section~6.2]{SEJ08}), i.e., the regex from our
running example, and handling of such cases remained an open problem.
 
Regular expressions with counters are also discussed
in~\cite{Hovland09,KilTuh07,GMN07}.
The automata with counters used in~\cite{Hovland09}, called FACs, correspond
closely, apart from our symbolic character predicates and transition
representation, to the class of CAs considered in our work.
A central result in~\cite{Hovland09} is that for \emph{counter-1-unambiguous}
regexes, the translation algorithm yields deterministic FACs and that checking
determinism of FACs can be done in polynomial time.
There are also works on regular expressions with counting that translate
deterministic regexes to CAs and work with different notions of
determinism~\cite{Gelade09_countingregex,Chen15_counting}.
The related work in~\cite{Hovland-Membership2012} studies membership in regexes
with counting.
None of these papers addresses the problem of determinising nondeterministic
CAs.

\section{Future Directions} 


Among future directions, we will consider optimisations of the current algorithm
by means of avoiding construction of unreachable parts of DCAs or by finding
efficient data structures, 
generalising the techniques used for monadic CAs to a larger class of CAs, 
and building a competitive pattern matching engine around the current algorithm. 
Since we believe that CAs have a  lot of potential as a general succinct
automata representation, we will work towards filling in efficient CA
counterparts of standard automata algorithms, such as Boolean operations,
minimisation, or emptiness test, that could also be used in other applications
than pattern matching, \mbox{such as verification and decision procedures of logics.}


\bibliographystyle{splncs}
\bibliography{literature}

\begin{thebibliography}{10}
\providecommand{\url}[1]{\texttt{#1}}
\providecommand{\urlprefix}{URL }
\providecommand{\doi}[1]{https://doi.org/#1}

\bibitem{abdulla_rautomata}
Abdulla, P.A., Kr\v{c}{\'{a}}l, P., Yi, W.: {R-Automata}. In: Proc. of
  {CONCUR}'08. LNCS, vol.~5201. Springer (2008)

\bibitem{finkel_fast}
Bardin, S., Finkel, A., Leroux, J., Petrucci, L.: {{FAST:} Acceleration from
  Theory to Practice}. {STTT}  \textbf{10}(5) (2008)

\bibitem{cikm15}
B\"orklund, E., Martens, W., Timm, T.: {Efficient Incremental Evaluation of
  Succinct Regular Expressions}. In: Proc. of {CIKM}'15. ACM (2015)

\bibitem{Chen15_counting}
Chen, H., Lu, P.: {Checking Determinism of Regular Expressions with Counting}.
  Information and Computation  \textbf{241} (2015)

\bibitem{cheng_efsm}
Cheng, K., Krishnakumar, A.S.: {Automatic Functional Test Generation Using the
  Extended Finite State Machine Model}. In: Proc. of DAC'93. {ACM} Press (1993)

\bibitem{SfaMinPOPL14}
D'Antoni, L., Veanes, M.: {Minimization of Symbolic Automata}. In: Proc. of
  POPL'14. ACM (2014)

\bibitem{DillHH91}
Dill, D.L., Hu, A.J., Wong-Toi, H.: Checking for language inclusion using
  simulation preorders. In: Proc. of CAV'91. LNCS, vol.~575. Springer (1992)

\bibitem{GMN07}
Gelade, W., Martens, W., Neven, F.: {Optimizing Schema Languages for XML:
  Numerical Constraints and Interleaving}. In: Proc. of ICDT'07. LNCS,
  vol.~4353. Springer (2007)

\bibitem{Gelade09_countingregex}
Gelade, W., Gyssens, M., Martens, W.: {Regular Expressions with Counting: Weak
  versus Strong Determinism}. In: Proc. of MFCS'09. LNCS, vol.~5734. Springer
  (2009)

\bibitem{5algos}
van Glabbeek, R., Ploeger, B.: {Five Determinisation Algorithms}. In: Proc. of
  CIAA'08. LNCS, vol.~5148. Springer (2008)

\bibitem{ultimate:ketchup}
Heizmann, M., Hoenicke, J., Podelski, A.: {Software Model Checking for People
  Who Love Automata}. In: Proc. of CAV'13. LNCS, vol.~8044. Springer (2013)

\bibitem{KlaEtAl:Mona}
Henriksen, J., Jensen, J., J{\o}rgensen, M., Klarlund, N., Paige, B., Rauhe,
  T., Sandholm, A.: Mona: Monadic second-order logic in practice. In: TACAS'95.
  LNCS, vol.~1019. Springer (1995)

\bibitem{Hovland09}
Hovland, D.: {Regular Expressions with Numerical Constraints and Automata with
  Counters}. In: Proc. of ICTAC'09. LNCS, vol.~5684. Springer (2009)

\bibitem{Hovland-Membership2012}
Hovland, D.: {The Membership Problem for Regular Expressions with Unordered
  Concatenation and Numerical Constraints}. In: Proc. of LATA'12. LNCS,
  vol.~7183. Springer (2012)

\bibitem{KilTuh07}
Kilpel\"ainen, P., Tuhkanen, R.: {One-unambiguity of Regular Expressions with
  Numeric Occurrence Indicators}. Information and Computation  \textbf{205}(6)
  (2007)

\bibitem{vata}
Leng{\'{a}}l, O., \v{S}im{\'{a}}\v{c}ek, J., Vojnar, T.: {VATA:} {A} library
  for efficient manipulation of non-deterministic tree automata. In: TACAS'12.
  LNCS, vol.~7214. Springer (2012)

\bibitem{snort}
{{M. Roesch et al.}}: {Snort: A Network Intrusion Detection and Prevention
  System}, {\url{http://www.snort.org}}

\bibitem{automata_library}
{Microsoft Automata library}: {Automata and Transducer Library for .NET},
  {\url{https://github.com/AutomataDotNet/Automata}}

\bibitem{owasp}
{OWASP Foundation and Checkmarx}: {Regular Expression Denial of Service:
  {ReDoS}} (2017)

\bibitem{regexlib}
RegExLib.com: {The Internet's First Regular Expression Library},
  {\url{http://regexlib.com/}}

\bibitem{bro}
{{Robin Sommer et al.}}: {The {Bro} Network Security Monitor},
  {\url{http://www.bro.org}}

\bibitem{shiple_efsm}
Shiple, T.R., Kukula, J.H., Ranjan, R.K.: {A Comparison of {Presburger} Engines
  for {EFSM} Reachability}. In: Proc. of CAV'98. LNCS, vol.~1427. Springer
  (1998)

\bibitem{SEJ08}
Smith, R., Estan, C., Jha, S.: {{XFA:} Faster Signature Matching with Extended
  Automata}. In: Proc. of SSP'08. IEEE (2008)

\bibitem{jha_extended}
Smith, R., Estan, C., Jha, S., Siahaan, I.: {Fast Signature Matching Using
  Extended Finite Automaton {(XFA)}}. In: Proc. of {ICISS}'08. LNCS, vol.~5352.
  Springer (2008)

\bibitem{Sperberg-McQueen-ExtendedFAWeb}
Sperberg-McQueen, M.: {Notes on Finite State Automata with Counters},
  \url{https://www.w3.org/XML/2004/05/msm-cfa.html}, accessed: 2018-08-08

\bibitem{sagan}
{{The Sagan team}}: {The {Sagan} Log Analysis Engine},
  \url{https://quadrantsec.com/sagan_log_analysis_engine/}

\bibitem{Thom68}
Thompson, K.: {Programming Techniques: Regular Expression Search Algorithm}.
  Communications of the ACM  \textbf{11}(6) (1968)

\bibitem{tacas18-appred}
\v{C}e\v{s}ka, M., Havlena, V., Hol\'{i}k, L., Leng\'{a}l, O., Vojnar, T.:
  {Approximate Reduction of Finite Automata for High-Speed Network Intrusion
  Detection}. In: TACAS'18. LNCS, vol. 10806. Springer (2018)

\bibitem{yang2010}
Yang, L., Karim, R., Ganapathy, V., Smith, R.: {Improving {NFA}-Based Signature
  Matching Using Ordered Binary Decision Diagrams}. In: RAID'10. LNCS,
  vol.~6307. Springer (2010)

\end{thebibliography}

%

\end{document}